\documentclass[a4paper,11pt]{article}
\pdfoutput=1 

\usepackage{jcappub} 

\usepackage{amsmath}
\usepackage{mathrsfs}
\usepackage[dvipsnames]{xcolor}
\usepackage{graphicx,amsfonts,comment,hyperref,float,times,subcaption}

\newcommand{\dd}{\; \mathrm{d}}
\newcommand{\isotope}[2]{${}^{\text{#2}}$#1}

\title{Direct Detection of Strongly Interacting Sub-GeV Dark Matter via Electron Recoils}


\author[a,b]{Timon Emken,}
\author[c]{Rouven Essig,}
\author[a,d]{Chris Kouvaris,}
\author[c]{and Mukul Sholapurkar}

\affiliation[a]{CP$^3$-Origins, University of Southern Denmark, Campusvej 55, DK-5230 Odense, Denmark}
\affiliation[b]{Chalmers University of Technology, Department of Physics, SE-412 96 G\"oteborg, Sweden}
\affiliation[c]{C.N. Yang Institute for Theoretical Physics, Stony Brook University, Stony Brook, NY 11794}
\affiliation[d]{CERN, Theoretical Physics Department, Geneva, Switzerland}

\emailAdd{emken@chalmers.se}
\emailAdd{rouven.essig@stonybrook.edu}
\emailAdd{kouvaris@cp3.sdu.dk}
\emailAdd{mukul.sholapurkar@stonybrook.edu}

\abstract{
We consider direct-detection searches for sub-GeV dark matter via electron scatterings in the presence of large interactions between dark and ordinary matter. Scatterings both on electrons and nuclei in the Earth's crust, atmosphere, and shielding material attenuate the expected local dark matter flux at a terrestrial detector, so that such experiments lose sensitivity to dark matter above some critical cross section. We study various models, including dark matter interacting with a heavy and ultralight dark photon, through an electric dipole moment, and exclusively with electrons.  For a dark-photon mediator and an electric dipole interaction, the dark matter-electron scattering cross-section is directly linked to the dark matter-nucleus cross section, and nuclear interactions typically dominate the attenuation process. We determine the exclusion bands for the different dark-matter models from several experiments --- SENSEI, CDMS-HVeV, XENON10, XENON100, and DarkSide-50 --- using a combination of Monte Carlo simulations and analytic estimates.  We also derive projected sensitivities for a detector located at different depths and for a range of exposures, and calculate the projected sensitivity for SENSEI at SNOLAB and DAMIC-M at Modane.  Finally, we discuss the reach to high cross sections and the modulation signature of a small balloon- and satellite-borne detector sensitive to electron recoils, such as a Skipper-CCD.  
Such a detector could potentially probe unconstrained parameter space at high cross sections for a sub-dominant component of dark matter interacting with a massive, but ultralight, dark photon.  
\\[0.2cm]

\noindent
\textit{Preprint: CERN-TH-2019-071, CP3-Origins-2019-18 DNRF90, YITP-SB-19-14}
}

\begin{document}
\maketitle
\flushbottom

\section{Introduction}
\label{s:introduction}

The nature of dark matter~(DM) is one of the biggest outstanding mysteries of particle physics. DM direct-detection experiments typically look for nuclear recoils induced by DM particles from the galactic halo~\cite{Goodman:1984dc,Wasserman:1986hh,Drukier:1986tm}. However, the nuclear recoils produced by DM with mass below about 100~MeV are currently challenging to detect as the recoil energies are below current detector thresholds. One of the ways around this problem is to look for electron recoils and other signals induced by sub-GeV DM in various materials~\cite{Essig:2011nj,Essig:2012yx,Graham:2012su,An:2014twa,Aprile:2014eoa,Lee:2015qva,Essig:2015cda,Hochberg:2015pha,Hochberg:2015fth,Aguilar-Arevalo:2016zop,Bloch:2016sjj,Cavoto:2016lqo,Derenzo:2016fse,Essig:2016crl,Hochberg:2016ntt,Hochberg:2016ajh,Hochberg:2016sqx,Kouvaris:2016afs,Budnik:2017sbu,Bunting:2017net,Cavoto:2017otc,Essig:2017kqs,Fichet:2017bng,Ibe:2017yqa,Knapen:2017ekk,Tiffenberg:2017aac,Hochberg:2017wce,Bringmann:2018cvk,Akerib:2018hck,Ema:2018bih}. Constraints now exist down to DM~masses of $\sim$$5-20$~MeV from xenon and argon detectors, and down to $\sim$500~keV from silicon detectors~\cite{Tiffenberg:2017aac,Romani:2017iwi,Crisler:2018gci, Agnese:2018col, Abramoff:2019dfb}. New experiments are already under development~\cite{Tiffenberg:2017aac,Settimo:2018qcm} and many ideas exist for future experiments~\cite{Battaglieri:2017aum}.

Direct-detection experiments are typically placed underground and have shielding to reduce cosmic-ray and radiogenic backgrounds. This,  however, limits the sensitivity of these detectors to DM that has large interactions with ordinary matter, which we will often refer to as ``strongly interacting DM.''\footnote{Here, ``strong'' refers to the size of the interaction between the DM and Standard Model (SM) particles, and not interactions described by Quantum Chromodynamics (QCD), i.e.~not the strong force.}  DM with large interactions can scatter off the nuclei and electrons in the Earth's crust, the atmosphere, or the shielding material of the experiment. These scatterings decelerate and deflect the DM particles, thereby reducing the DM flux at the detector. Above a critical cross section, the~DM particles scatter so often before reaching the detector that they are unable to cause a detectable signal~\cite{Goodman:1984dc,Starkman:1990nj}. Direct-detection experiments can therefore probe only a band of DM-electron cross section values: below the lower limit, the number of events produced in the detector are too few to detect due to the weak interaction strength. Above the upper limit, the terrestrial effects described above diminishes the detector's sensitivity.

The terrestrial effect for conventional~DM searches via nuclear recoils has been studied earlier, both analytically in~\cite{Starkman:1990nj,Kouvaris:2014lpa,Kavanagh:2016pyr,Davis:2017noy,Kavanagh:2017cru,Bramante:2018qbc,Armengaud:2019kfj,Laletin:2019qca}, or via Monte Carlo~(MC) simulations. In the latter case, MC simulations were used to predict diurnal signal modulations~\cite{Collar:1993ss,Hasenbalg:1997hs,Emken:2017qmp}, and to describe the stopping effect of a detector's overburden, first in~\cite{Zaharijas:2004jv} and then later in~\cite{Mahdawi:2017cxz,Emken:2018run}. Furthermore, more general velocity-dependent DM-nucleus interactions, including milli-charged~DM, were studied in~\cite{Mahdawi:2018euy}. In the case of electron recoil experiments, the terrestrial effect was studied with MC simulations in~\cite{Emken:2017erx}, where the constraints were derived from a simple speed criterion, and only a heavy dark-photon mediator with momentum-independent interactions was considered.       

In this paper, we discuss in detail the case where DM interacts with a dark photon that is kinetically mixed with the SM hypercharge gauge boson~\cite{Galison:1983pa,Holdom:1985ag}, the scenario in which the DM-SM interaction is through an electric dipole moment, and the possibility that DM interacts only with electrons and not quarks and nuclei. For the dark-photon and electron-only interactions, we investigate both the case that the mediator mass is heavy (much larger than a keV, the typical momentum transfer when the DM scatters off electrons in atoms or semiconductors) and ultralight (well below a keV).  For a dark-photon mediator, the ultralight case is especially interesting, since it is usually neglected  in past studies of terrestrial effects on direct-detection experiments.  Moreover, constraints on such DM from other, non-direct-detection probes, especially from colliders, are significantly weaker than when the mediator mass is heavy.  We note that in the limit of zero dark-photon mass, the DM is milli-charged.  

For these models, we calculate the terrestrial effect using a combination of MC techniques (where possible) and analytic approximations, and calculate the upper limit of the DM-electron scattering cross section that can be probed by direct-detection experiments sensitive to electron recoils. We consider several stopping processes, including elastic scatterings of DM with the nuclei and electrons in the medium, as well as inelastic scatterings like ionization of electrons. We present the band of cross section that is probed by the data taken by the experiments XENON10~\cite{Angle:2011th,Essig:2017kqs}, XENON100~\cite{Aprile:2016wwo,Essig:2017kqs}, SENSEI~\cite{Crisler:2018gci,Abramoff:2019dfb}, DarkSide-50~\cite{Agnes:2018oej}, and SuperCDMS~\cite{Agnese:2018col}. We also give projections for future experiments, and how the upper boundaries change with the position and exposure of the experiment. Finally, we consider the cross sections that could be probed with a new direct-detection experiment placed on a balloon or a satellite. 

Dark matter with a light or massless dark photon mediator, including milli-charged DM, is constrained by several astrophysical and cosmological observations. These constraints are based on, for example, the Cosmic Microwave Background~(CMB)~\cite{Chen:2002yh,Dvorkin:2013cea,Gluscevic:2017ywp,McDermott:2010pa,Slatyer:2015jla,Dolgov:2013una,Dubovsky:2003yn,Xu:2018efh,Ali-Haimoud:2015pwa,Boddy:2018wzy}, Big Bang Nucleosynthesis~(BBN)~\cite{Vogel:2013raa,Davidson:2000hf,Creque-Sarbinowski:2019mcm,Boehm:2013jpa}, Galactic center gas clouds~\cite{Bhoonah:2018gjb}, or Supernova 1987A~\cite{Chang:2018rso,Davidson:2000hf}. Colliders and beam-dump experiments have also set bounds on milli-charged dark matter~\cite{Davidson:2000hf,Prinz:1998ua,Magill:2018tbb}. However, for a strongly interacting subdominant component of DM, there currently seems to be an open window between the astrophysical, cosmological, and accelerator-based constraints, and direct detection constraints by surface and underground experiments. This provides motivation to investigate the prospect of placing a detector sensitive to electron recoils, e.g. a Skipper-CCD~\cite{Tiffenberg:2017aac}, aboard a balloon or a satellite. At such high altitudes, the DM flux is attenuated only very weakly and much stronger interactions can be probed directly~\cite{snowden1990search}. 
Even if the backgrounds from cosmic rays are large, the expected signal is large as well and very distinctive, showing e.g.~a strong modulation due to the Earth's shadowing effect~\cite{10150/186654}.

The outline of the rest of this paper is as follows. In Sec.~\ref{s:model}, we give the details of the DM model and interactions considered in this paper. In Sec.~\ref{s:stopping}, we give the equations used to calculate the stopping power of~DM interactions with both nuclei and electrons. In Sec.~\ref{s:results}, we show the band of DM-electron cross sections that are excluded by existing data from XENON10, XENON100, SENSEI, DarkSide-50, and SuperCDMS.  This chapter also includes the projections for future experiments, including satellite and balloon-borne 
experiments. Sec.~\ref{s:conclusions} summarizes our conclusions.  Three appendices provide additional details on the calculations.  \\[0.1cm] 

Along with this paper, we release version 1.1 of the~\textsc{DaMaSCUS-CRUST} tool~\cite{Emken2018a}, which was originally made available alongside~\cite{Emken:2018run}. This code was used to generate all~MC results reported in this paper. It introduces the option of light mediators, electric dipole interactions, DM-electron scattering experiments, as well as the rare event technique of Geometric Importance Splitting, which is described in more detail in Appendix~\ref{app:splitting}.  The code is publicly available on \href{https://github.com/temken/DaMaSCUS-CRUST}{GitHub}, an archived version can also be found under~\href{https://doi.org/10.5281/zenodo.2846401}{[DOI:10.5281/zenodo.2846401]}.

\section{Dark Matter Models and Interactions}
\label{s:model}

We discuss three DM models in this paper: DM coupled to a dark photon, DM interacting with the SM through a electric dipole moment, and DM that interacts exclusively (or dominantly) with electrons only.  The first two models allow for DM interactions with both electrons and nuclei, and the latter often dominate the terrestrial stopping effects.  
We describe in detail in this section the dark-photon interaction, which we use to set up the notation and to explain important features of the direct-detection cross sections, before briefly mentioning the other two models at the end.  

\subsection*{Dark Matter Interacting with a Dark Photon}
A simple model of the dark sector is realized by extending the SM via a fermionic DM field, $\chi$, of mass~$m_{\chi}$, which is charged under a new broken Abelian gauge symmetry $U(1)_D$.  We allow for kinetic mixing of the dark photon to the $U(1)_Y$ gauge boson, where~$Y$ is the hypercharge of the SM~\cite{Galison:1983pa,Holdom:1985ag}. At low energies, the dominant kinetic mixing is between the dark photon and the SM~photon, and the effective Lagrangian of the dark sector is given by
\begin{subequations}
\begin{align}
	\mathscr{L}_D &= \bar{\chi} (i\gamma^{\mu}D_\mu-m_{\chi}) \chi  + \frac{1}{4}F'_{\mu \nu}F'^{\mu \nu} + m_{A^\prime}^{2} A'_{\mu}A'^{\mu}+ \frac{\varepsilon}{2} F_{\mu \nu}F'^{\mu \nu}\, ,
	\intertext{with the covariant derivative}
	D_\mu&=\partial_\mu - i g_D A'_\mu\, ,
	\intertext{and the field strength tensors}
	F_{\mu \nu}&=\partial_\mu A_\nu -\partial_\nu A_\mu \, ,\quad F^\prime_{\mu \nu}=\partial_\mu A^\prime_\nu -\partial_\nu A^\prime_\mu \, .
\end{align}
\end{subequations}
This Lagrangian introduces a dark photon field~$A'$ of mass~$m_{A^\prime}$, its kinetic mixing parameter~$\varepsilon$ with the photon field, as well as the gauge coupling~$g_D$ of the $U(1)_D$~gauge group. The mass term of the dark photon breaks gauge invariance, and the new symmetry is assumed to be broken, possibly via a dark Higgs mechanism, which we do not 
specify further. 

The mixing between the $U(1)_{\rm em}$ and~$U(1)_D$ gauge bosons generates an interaction between the dark photon and charged fermions of the~SM~\cite{Kaplinghat:2013yxa}. One can parametrize the couplings of the mediator to protons~($p$) and electrons~($e$) as\footnote{The sub-dominant mixing with the~$Z$ boson induces an additional, but very weak coupling to neutrons, and is not relevant in this paper.},
\begin{align}
	\mathscr{L}_{\rm int} = e \varepsilon A^\prime_\mu \left(\bar{p}\gamma^\mu p- \bar{e}\gamma^\mu e\right)\, .
\end{align}
The differential scattering cross section between DM and a nucleus ${}^{A}_{Z}N$ in this model is given by
\begin{align}
		\frac{\dd \sigma_{N}}{\dd q^2} &= \frac{4\pi \alpha\alpha_D\epsilon^2}{(q^2+m_{A^\prime}^2)^2}\frac{1}{v^2}F_N(q)^2 Z^2\, ,
\end{align}
where $\alpha\equiv e^2/(4\pi)~(\alpha_D\equiv g_D^2/(4\pi))$ is the fine structure constants of the SM~(dark) sector, $q$ is the momentum transfer of the scattering, $v$ is the relative speed between nucleus and DM particle, and $F_N(q)$ is the nuclear form factor, which accounts for the loss of coherence for large momentum transfers. For DM-electron scattering the corresponding differential cross section is
\begin{align}
	\frac{\dd \sigma_{e}}{\dd q^2} &= \frac{4\pi \alpha\alpha_D\epsilon^2}{(q^2+m_{A^\prime}^2)^2}\frac{1}{v^2}\, .
\end{align}
Furthermore, we define reference cross sections
\begin{align}
\bar{\sigma}_p&\equiv \frac{16\pi \alpha\alpha_D\epsilon^2 \mu_{\chi p}^2}{(q_{\rm ref}^2+m_{A^\prime}^2)^2}
\intertext{for DM-proton scatterings, and}
	\bar{\sigma}_e&\equiv \frac{16\pi \alpha\alpha_D\epsilon^2 \mu_{\chi e}^2}{(q_{\rm ref}^2+m_{A^\prime}^2)^2} 
\end{align}
for DM-electron interactions. Throughout this paper $\mu_{ij}$ denotes the reduced mass of particle~$i$ and~$j$. Note that the reference cross sections depend on an arbitrary reference momentum transfer $q_{\rm ref}$, unless we consider the heavy mediator limit. We choose $q_{\rm ref}$ to be $\alpha m_e$, the typical momentum transfer in DM-electron collisions for noble-liquid and semiconductor targets. 
The ratio of the two reference cross sections does not depend on this arbitrary choice and is given by 
\begin{align}
\frac{\bar{\sigma}_p}{\bar{\sigma}_e} = \left(\frac{\mu_{\chi p}}{\mu_{\chi e}}\right)^2\, .\label{eq:sigma ratio}
\end{align}
This naturally leads to a hierarchy between the two cross sections. For MeV-scale DM masses and heavier, we find $\bar{\sigma}_p\gg \bar{\sigma}_e$. This is why DM-nucleus scatterings in the Earth's crust and atmosphere can easily become non-negligible for DM-electron scattering experiments~\cite{Lee:2015qva,Emken:2017erx}. 

It is useful to write the differential cross sections in terms of~$\bar{\sigma}_p$ and~$\bar{\sigma}_e$,  
\begin{align}
	\frac{\dd \sigma_{N}}{\dd q^2} &= \frac{\bar{\sigma}_p}{4\mu_{\chi p}^2v^2}F_{\rm DM}(q)^2\,F_N(q)^2\, Z^2\, ,\label{eq:dsdq2 nucleus}\\
	\frac{\dd \sigma_{e}}{\dd q^2} &=\frac{\bar{\sigma}_e}{4\mu_{\chi e}^2v^2}F_{\rm DM}(q)^2\, , 
	\intertext{with the DM form factor}
	F_{\rm DM}(q) &= \frac{q_{\rm ref}^2+m_{A^\prime}^2}{q^2+m_{A^\prime}^2}\, ,\label{eq: DM form factor}
\end{align}
which parametrizes the $q$ dependence.

In this model, the DM field~$\chi$ couples to the electric charge of the SM~fields via kinetic mixing of the mediator with the SM~photon. The above cross sections assume free nuclei and electrons, whereas the charges in an overall neutral medium are screened by the surrounding charges on large distances. Assuming a simple screened Coulomb potential $\frac{Ze}{r}e^{-r/a}$ for the scattering on nuclei, we can account for this effect by rescaling the nuclear charge,
\begin{align}
	Z \rightarrow Z_{\rm eff} &= F_A(q) \times Z\, ,
\intertext{where we introduced the atomic form factor~$F_A(q)$, which satisfies $\lim\limits_{q\rightarrow \infty}F_A(q)=1$ and $F_A(0)=0$. It can be written in a compact form,}
	F_A(q) &= \frac{a^2 q^2}{1+a^2q^2}\, , \label{eq: atomic form factor}
	\intertext{with the Thomas-Fermi radius}
	a&= \frac{1}{4}\left(\frac{9\pi^2}{2Z}\right)^{1/3}a_0\approx \frac{0.89}{Z^{1/3}}a_0\, ,
\end{align}
where $a_0$ is the Bohr radius. The atomic form factor in Eq.~\eqref{eq: atomic form factor} approximates the corresponding result in the Thomas-Fermi model~\cite{Schiff:1951zza,Tsai:1973py} and decreases the effective nucleus charge on large distances, i.e.~for small $q$, as the the nucleus gets more and more screened by electrons.

The total DM-nucleus scattering cross section is obtained by integrating Eq.~\eqref{eq:dsdq2 nucleus} including the atomic form factor,
\begin{align}
	\sigma^{\rm tot}_{N} &= \int\limits_0^{q^2_{\rm max}}\dd q^2\;\frac{\dd \sigma_{N}}{\dd q^2}F_A(q)^2\,, \label{eq:totalCS}
\end{align}
with $q_{\rm max}^2 = 4\mu_{\chi N}^2v^2$.  
Since we are interested in low-mass~DM and low momentum transfers, we can neglect the nuclear form factor by approximating~$F_N(q)\approx 1$. In this case, the integration in Eq.~\eqref{eq:totalCS} can be performed analytically.  
It is convenient to consider the two cases of ultralight and heavy dark photons, expressed by the DM form factor
\begin{align}\label{eq:FDM-A'}
	F_{\rm DM}(q) =
	\begin{cases}
		1\, ,\;&\text{for }m_{A^\prime}^2\gg q^2_{\rm max}\, ,\\
		\left(\frac{q_{\rm ref}}{q}\right)^2\, ,\;&\text{for }m_{A^\prime}^2\ll q^2_{\rm max}\, .
	\end{cases}
\end{align}
The total cross sections for these two cases are 
\begin{align}
	\sigma^{\rm tot}_{N} &=\bar{\sigma}_p\left(\frac{\mu_{\chi N}}{\mu_{\chi p}}\right)^2Z^2\times
	\begin{cases}
		\left(1+\frac{1}{1+a^2 q^2_{\rm max}}-\frac{2}{a^2q^2_{\rm max}}\log(1+a^2 q^2_{\rm max})\right)\, , &\text{for }F_{\rm DM}(q)=1\, ,\\[2ex]
		\frac{a^4 q_{\rm ref}^4}{(1+a^2q^2_{\rm max})}\, , &\text{for }F_{\rm DM}(q)= \left(\frac{q_{\rm ref}}{q}\right)^2\, .
	\end{cases}
\end{align} 

The common pre-factor in isolation is the usual result for contact interactions without screening. The case-specific factor depends on the screening length~$a$. In the first case, it is $\sim$1 for $m_{\chi}>100$~MeV; in other words, charge screening has a negligible impact on heavier DM. In the case of an ultralight dark photon, we obtain a Rutherford-type cross section with the typical~$\sim\frac{1}{q^4}$ behavior diverging in the IR. However, the atomic form factor mimics a finite mediator mass, which renders the total cross section finite. 

\subsection*{Dark Matter Interacting with an Electric Dipole Moment}
We consider also an electric dipole moment interaction~\cite{Sigurdson:2004zp} of the form 
\begin{equation}
\mathscr{L}_{\rm int} \supset \frac{1}{\Lambda}\overline{\chi}\sigma^{\mu\nu} \gamma^5  \chi F_{\mu\nu}\, , \quad\text{with }\sigma_{\mu \nu}= \frac{i}{2} [\gamma_{\mu},\gamma_{\nu}]\, ,
\end{equation}
where $\Lambda$ is the scale at which the electric dipole moment is generated.  
This type of interaction can conveniently be embedded into the previous framework of the dark photon model by setting the 
DM~form factor to
\begin{equation}\label{eq:FDM-EDM}
F_{\rm DM}(q) = \frac{q_{\rm ref}}{q}\, .
\end{equation} 
The total DM-nucleus scattering cross section is given by 
\begin{equation}
	\sigma^{\rm tot}_{N} =\bar{\sigma}_p\left(\frac{\mu_{\chi N}}{\mu_{\chi p}}\right)^2Z^2\times
\left[ 
\frac{q_{\rm ref}^2}{q^2_{\rm max}}\left(\log(1+a^2q^2_{\rm max})-\frac{a^2 q^2_{\rm max}}{1+a^2 q^2_{\rm max}}\right)
\right]\,.
\end{equation} 
Despite the fact that this interaction does not arise in the dark photon model, and this form factor cannot be obtained directly from Eq.~\eqref{eq: DM form factor}, electric dipole interactions can easily be included and studied this way, and the analysis is very similar to the dark photon case.

\subsection*{Dark Matter Interacting with Electrons Only}
Finally, we consider the phenomenological case in which the DM interacts only (or dominantly) with electrons rather than the nuclei.  In other words, we assume that $\bar{\sigma}_p \approx 0$. We note that DM-nucleus interactions often get generated at the loop-level even in the absence of tree-level DM-quark interactions.  However, in leptophilic models with either a pseudoscalar or an axial vector DM-lepton interaction, no DM-nucleus interactions will be generated at the loop-level, and the~DM remains `truly' leptophilic~\cite{Kopp:2009et}.

\section{Signal attenuation by the Earth's Atmosphere and Overburden}
\label{s:stopping}

The two main target materials used to search for DM-electron scatterings so far are noble liquids and semiconductors. For atomic targets such as liquid xenon or argon, the ionization rate of an electron to final  energy $E_e$ in a detector of~$N_T$ target atoms is given by~\cite{Essig:2011nj,Essig:2012yx,Essig:2017kqs}
\begin{subequations}
\label{noble}
\begin{align}
	\frac{\dd R_{\rm ion}}{\dd E_e} & = N_T \frac{\rho_{\chi}}{m_{\chi}}\sum_{nl} \frac{\dd \langle \sigma_{\rm ion}^{nl}v\rangle}{\dd E_e}\, ,
	\intertext{where we substitute the speed-averaged differential ionization cross section,}
	\frac{\dd \langle \sigma_{\rm ion}^{nl}v\rangle}{\dd E_e}&=\frac{\overline{\sigma}_e}{8\mu_{\chi e}^2E_e}\int\dd q\,\left|F_{\rm DM}(q)\right|^2\left|f_{\rm ion}^{nl}(k,q)\right|^2\eta(v_{\rm min}(q,E_e))\, .
\end{align}
\end{subequations}
Here, we used the local DM energy density $\rho_\chi=0.3\text{ GeV/cm}^3$~\cite{Bovy:2012tw}, $(n,l)$ identifies the atomic shells, $f(v)$ is the local speed distribution of the DM at the detector, and $f_{\rm ion}^{nl}(k,q)$ is the ionization form factor for a given shell, where $k = \sqrt{2m_e E_e}$ is the electron's final momentum. If DM stopping in the overburden is negligible, $f(v)$ is simply the DM halo speed distribution~$f_{\rm halo}(v)$. The standard choice in the context of direct detection is a truncated Maxwell-Boltzmann distribution, transformed into the Earth's rest frame,
\begin{align}
	f_{\rm halo}(\mathbf{v}) &= \frac{1}{N_{\rm esc}(\sqrt{\pi}v_0)^3}e^{-\frac{(\mathbf{v}+\mathbf{v}_\oplus)^2}{v_0^2}}\Theta(v_{\rm esc}-|\mathbf{v}+\mathbf{v}_\oplus|)\, ,
	\intertext{with the normalization constant~$N_{\rm esc}={\rm erf}\left( \frac{v_{\rm esc}}{v_0}\right)-\frac{2}{\sqrt{\pi}}\frac{v_{\rm esc}}{v_0}\exp \left(-\frac{v_{\rm esc}^2}{v_0^2} \right)$, the Earth's velocity in the galactic rest frame~$\mathbf{v}_\oplus$, the velocity dispersion~$v_0=220$~km/sec~\cite{Kerr:1986hz}, and the galactic escape velocity~$v_{\rm esc}=544$~km/sec~\cite{Smith:2006ym}. The speed distribution~$f_{\rm halo}(v)$ is obtained by integrating out the directional dependence of the full velocity distribution,}
	f_{\rm halo}(v) &= \int \dd \Omega\; v^2 f_{\rm halo}(\mathbf{v})\, ,
\end{align}
and will be used to sample initial conditions for the~MC simulations. The speed of the Earth in the galactic rest frame varies throughout the year. We do not specify the exact time of the experiments and set~$v_\oplus\approx 240$~km/sec.

The minimum speed required by a DM particle to change a bound electron's energy by~$\Delta E_e$ through a momentum transfer of~$q$ is
\begin{align}
	v_{\rm min}(q,E_e) = \frac{\Delta E_e}{q}+\frac{q}{2m_\chi}\, .
\end{align}
In the case of ionization of an electron of atomic shell~$(n,l)$, the transferred energy is~$\Delta E_e=|E_{\rm b}^{nl}|+E_e$, where $E_{\rm b}^{nl}$ is the binding energy of the corresponding atomic state.

For semiconductor targets, such as silicon or germanium crystals, the differential excitation rate in terms of the total deposited energy~$E_e$ is derived in~\cite{Essig:2011nj,Essig:2015cda}, and given by
\begin{align}\label{semiconductor}
	\frac{\dd R_{\rm cryst}}{\dd E_e} &= \frac{\rho_\chi}{m_\chi} \frac{M_{\rm target}}{M_{\rm cell}}\frac{\overline{\sigma}_e\alpha m_e^2}{\mu_{\chi e}^2}\int \dd q\;\frac{1}{q^2}\eta(v_{\rm min}(q,E_e))F_{\rm DM}(q)^2 \left| f_{\rm crystal}(q,E_e)\right|^2\, .
\end{align}
The cell mass for a silicon (germanium) crystal is $M_{\rm cell}=2m_{\rm Si(Ge)}$. The crystal form factor, $f_{\rm crystal}(q,E_e)$, encompasses all the information about the semiconductor's electronic band structure. It was computed and tabulated with the QEdark tool~\cite{Essig:2015cda}\footnote{Available at~\url{http://ddldm.physics.sunysb.edu/ddlDM/}.}.

In both Eq.~\eqref{noble} and~\eqref{semiconductor}, the local DM speed distribution~$f(v)$ at the detector enters via the function $\eta(v_{\rm min})$, given by
\begin{align}\label{eta}
	\eta(v_{\rm min})\equiv \int\limits_{v>v_{\rm min}}\dd v\, \frac{f(v)}{v}\, .
\end{align}
Pre-detection scatterings on matter will deform the speed distribution and decrease the underground DM~flux, leading to a signal attenuation. As a DM particle traverses a medium it loses energy and gets deflected via elastic collisions with nuclei and electrons, as well as inelastic scatterings such as ionization of atoms or excitations in crystals. The purpose of the following subsections is to estimate and compare the different stopping processes, and to show how the direct-detection rates and the upper limits on the cross sections are calculated incorporating the stopping effect.  

The so-called \textit{stopping power} is a local analytic measure of the average energy loss of a particle passing through and interacting with matter. Each of the above mentioned scattering processes contribute to the overall stopping power of a DM~particle,
\begin{align}
	S_{\rm tot}(\mathbf{x},E_\chi)\equiv -\left\langle\frac{\dd E_\chi}{\dd x}\right\rangle = S_n(\mathbf{x},E_\chi) + S_e(\mathbf{x},E_\chi) +S_a(\mathbf{x},E_\chi)\, .\label{eq: stopping power total}
\end{align}

The contribution to the overall stopping power due to elastic scatterings on nuclei, ionization and excitation of bound electrons, and elastic scatterings on atoms via an electron interaction are denoted as $S_n$, $S_e$, and $S_a$ respectively. In the following, we treat the three stopping powers separately and calculate the critical cross sections separately for each. We then set the final critical cross section to be the lowest of these three. This approximation is almost exact in the case when one of the stopping powers dominates over the other two. We find that if the incoming DM~particles can scatter elastically on nuclei, $S_n$ dominates over $S_e$ and $S_a$, and the upper boundary of the direct detection constraint band is set by $S_n$. In the case of electron-only interactions, the critical cross section is set to be the minimum of the critical cross sections based on atomic scattering ($S_a$) and ionization ($S_e$). Generally, it is much harder to describe these processes in a medium such as rock. We will find analytic estimates of this effect and the corresponding bounds.

\subsection{Nuclear Stopping Power and MC Simulations}
In this subsection, we estimate the stopping effect of elastic collisions of DM particles with the nuclei in the medium. In the previous section, we derived total cross sections for the scattering, which we can now use to define the mean free path of a DM particle inside a medium of resting targets, 
\begin{align}
	\lambda^{-1}(\mathbf{x},v) = \sum_i n_i(\mathbf{x}) \sigma_{i}^{\rm tot}\, .
\end{align}
In general, $\lambda$ is a local quantity and may also depend on the DM particle's speed $v$. The index $i$ runs over all constituent particles abundant in that medium with number density $n_i$. The mean free path is a measure of how often a given DM particle is expected to scatter when traveling a given distance underground. However, it tells us nothing about how efficiently these scatterings slow it down, which strongly depends on both the DM mass as well as the mediator mass. For heavy mediators and GeV scale DM, a single elastic scattering with a nucleus can suffice to significantly decelerate or even stop the DM particle. In contrast, for ultralight mediators forward scattering might be favored to an extent that a single scatter has very little impact on the trajectory of the DM particle.  A good measure which takes this into account is the stopping power, which we introduced previously. It takes into account both the likelihood to scatter in the first place and the typical energy loss per scattering. 

The nuclear stopping power via elastic collisions is given by
\begin{align}\label{eq:stopping}
	S_n(\mathbf{x},E_\chi) = \sum_i n_i(\mathbf{x}) \int\limits_{0}^{E_{R_i}^{\rm max}}\dd E_R \; E_R\frac{\dd \sigma_{i}}{\dd E_R}\, .
\end{align}
Just like $\lambda$, the nuclear stopping power $S_n$ is a local quantity, which generally depends on the energy of the DM particle. Here we used the differential cross section in terms of the recoil energy,
\begin{align}
	\frac{\dd \sigma_{i}}{\dd E_R} = 2 m_i\left.\frac{\dd \sigma_{i}}{\dd q^2}\right|_{q^2 = 2 m_i E_R}\, , 
\end{align}
where $m_i$ is the mass of the $i^{\rm th}$ target. It might be interesting to note that the integrated nuclear stopping power of the entire atmosphere is approximately equivalent to about 5 meters of rock/concrete or 2 meters of lead shielding.

This energy loss affects the rate of events produced at the direct detection experiments as the DM particles traveling through the medium get scattered off the nuclei before reaching the detector, and their speed distribution changes. This was investigated using the analytic approach of the stopping power for contact interactions~\cite{Starkman:1990nj,Kouvaris:2014lpa,Kouvaris:2015laa,Davis:2017noy,Kavanagh:2017cru}, dipole interactions~\cite{Sigurdson:2004zp}, and light mediators~\cite{Mahdawi:2018euy}. In contrast, we use MC techniques to simulate the nuclear scatterings and precisely calculate the change in the underground distribution of DM particles. Similar MC~simulations have been applied in this context previously~\cite{Zaharijas:2004jv,Erickcek:2007jv,Emken:2017erx,Mahdawi:2017cxz,Emken:2018run,Mahdawi:2018euy}.  

\paragraph{Monte Carlo Simulations}

For the MC treatment of nuclear stopping, we build upon the \textsc{DaMaSCUS-Crust} tool which was published in~\cite{Emken:2018run,Emken2018a}. Three major extension have been implemented in an updated version to obtain the MC results presented in this paper.
\begin{enumerate}
	\item In addition to contact interactions, we study more general interactions, most importantly DM-nucleus couplings mediated by an ultralight dark photon. This alters the scattering kinematics considerably as we will discuss further below.
	\item In the simulations, a detectable DM particle reaching the detector depth is typically a rare event; so rare, that brute-force MC simulations become inefficient, and in some cases practically impossible. For example, in the collision of a GeV-scale DM particle, which interacts via an ultralight dark photon, forward scattering is highly favored. Since the relative loss of kinetic energy is tiny, it requires a great number of scatterings to attenuate the detectable DM flux, which in turn increases the simulation time. Importance sampling (IS), as used in~\cite{Mahdawi:2017cxz,Emken:2018run}, is not applicable in this case. Instead, we implement adaptive geometric importance splitting~(GIS), a proven and reliable MC~method for rare-event simulation. In Appendix~\ref{app:splitting} we comment on the problems of~IS in this context and discuss~GIS in detail.
	\item  On the data-analysis side, we focus exclusively on DM-electron scattering experiments~\cite{Essig:2011nj}. With ionization of target atoms and excitations of electrons in crystals being the primary signal, the computation of event rates requires knowledge of the ionization and crystal form factors found in~\cite{Essig:2012yx,Essig:2017kqs,Essig:2015cda}.  We refer to Appendix~\ref{app:experiments} for more details about the data.
\end{enumerate}

\begin{figure}[!t]
\vspace{0mm}
\includegraphics[width=0.50\textwidth]{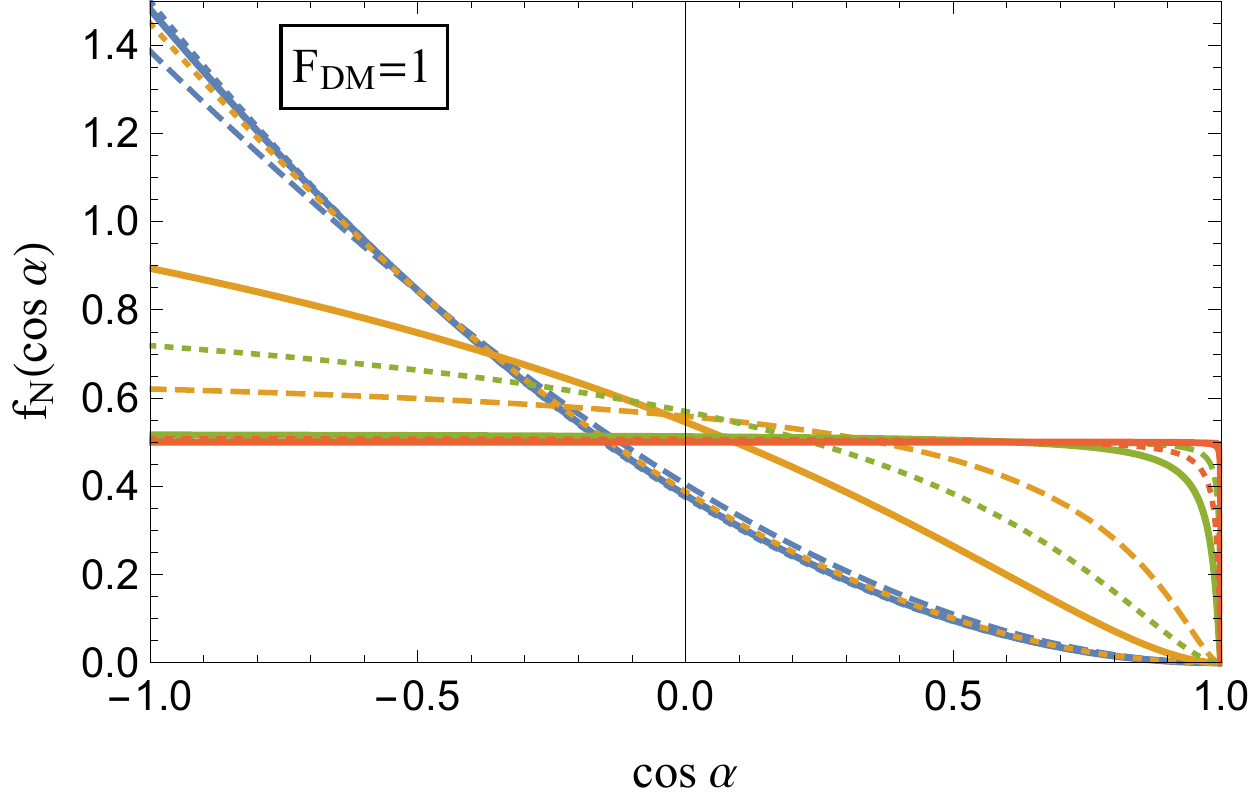}
~\includegraphics[width=0.50\textwidth]{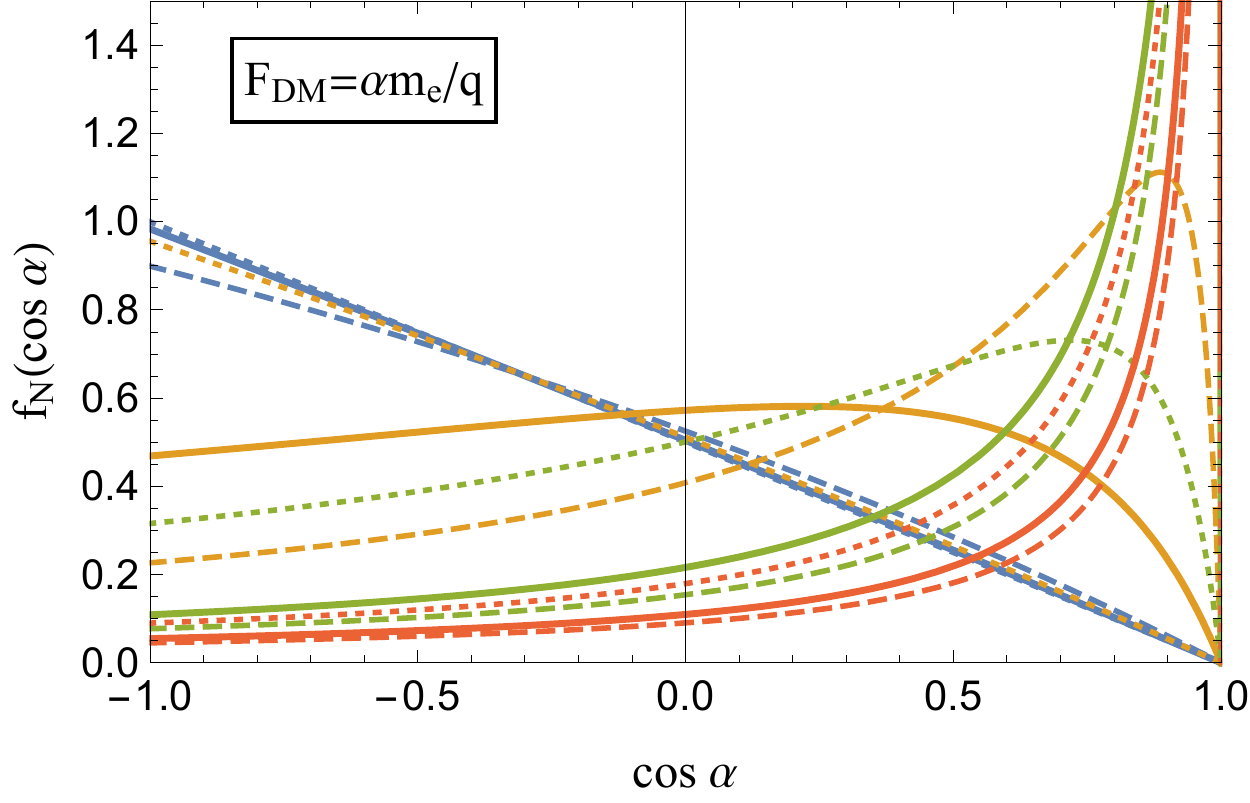}
\vspace{-7mm}
\begin{center}
	\includegraphics[width=0.75\textwidth]{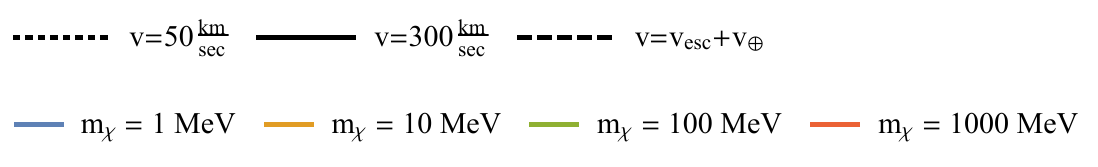}
\end{center}
\vspace{0mm}
\begin{minipage}[t]{0.5\textwidth}
\mbox{}\\[-\baselineskip]
\includegraphics[width=\textwidth]{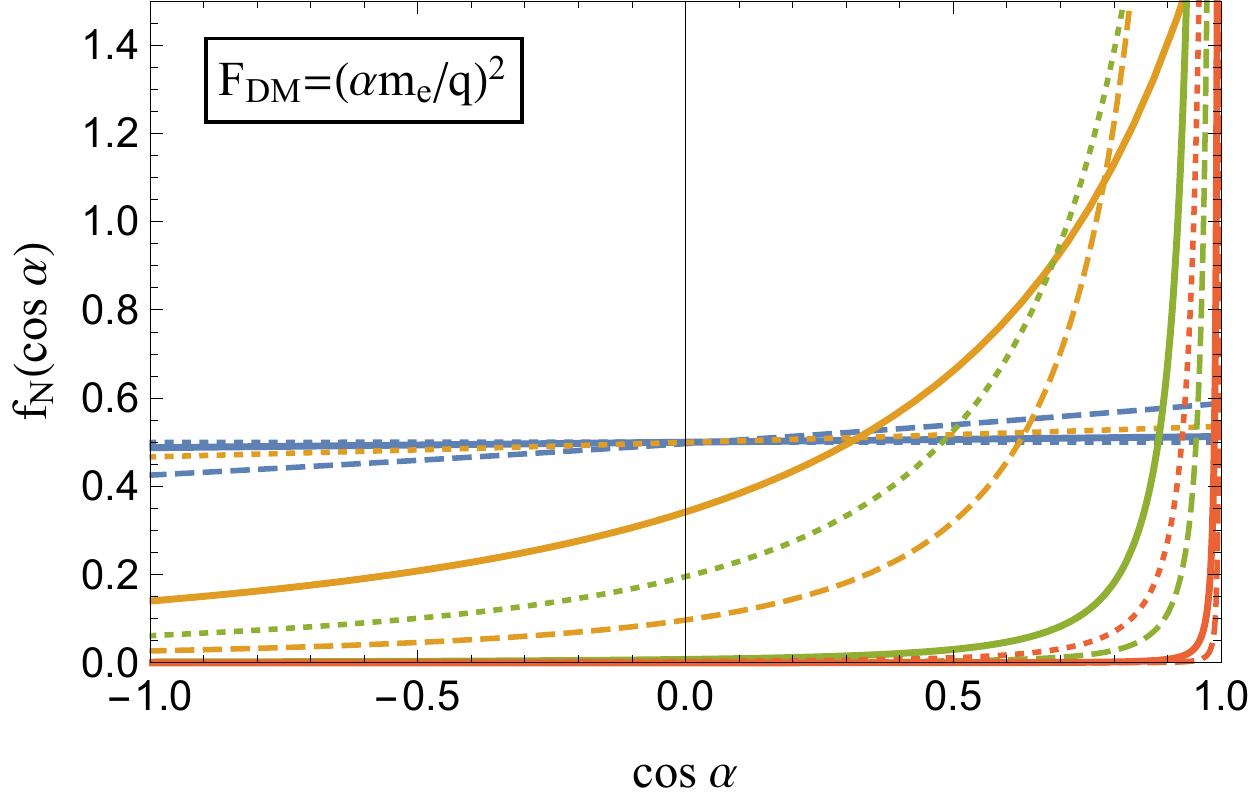}
\end{minipage}
\hfill
\begin{minipage}[t]{0.45\textwidth}
\mbox{}\\[-\baselineskip]
\vspace{-23.5pt}
\caption{Probability density functions~$f_N(\cos\alpha)$ of the scattering angle~$\alpha$ (or rather its cosine) for different DM masses and speeds in the case of contact, electric dipole, and long range interactions. The example target is chosen to be an iron nucleus~$\left({}^{56}_{26}\text{Fe}\right)$.
}
\label{fig:pdfcosa}
\end{minipage}
\end{figure}

The main difference to previous works is the consideration of light mediators, which changes the scattering kinematics. The distribution function for the cosine of the scattering angle $\alpha$ with a target nucleus $N$ can be inferred from the differential cross sections in Eq.~\eqref{eq:dsdq2 nucleus},
\begin{align}
	f_N(\cos \alpha) &=
	 \frac{1}{\sigma_N^{\rm tot}}\frac{\dd \sigma_N}{\dd \cos\alpha}=\frac{1}{2}\frac{q^2_{\rm max}}{\sigma_{N}^{\rm tot}}\frac{\dd \sigma_N}{\dd q^2}\, ,
\end{align}
where we substitute $q^2 = 2\mu_{\chi N}^2v^2(1-\cos\alpha)$. Hence, the scattering angle distributions are determined by the two form factors $F_A(q)$ and $F_{\rm DM}(q)$. We find
\begin{align}
	f_N(\cos \alpha) &= \frac{1}{2}\times
	\begin{cases}
		\frac{x^3}{4}\frac{1+x}{x(2+x)-2(1+x)\log(1+x)}\frac{(1-\cos\alpha)^2}{\left(1+\frac{x}{2}(1-\cos\alpha)\right)^2}\, ,\quad &\text{for }F_{\rm DM}(q)=1\, ,\\[2ex]
		\frac{x^2}{2}\frac{1+x}{(1+x)\log(1+x)-x}\frac{(1-\cos\alpha)^2}{\left(1+\frac{x}{2}(1-\cos\alpha)\right)^2}\, ,\quad &\text{for }F_{\rm DM}(q)\sim\frac{1}{q}\, ,\\[2ex]
	 \frac{1+x}{\left(1+\frac{x}{2}(1-\cos\alpha)\right)^2}\, ,\quad &\text{for }F_{\rm DM}(q)\sim\frac{1}{q^2}\, ,
	\end{cases}
\end{align}
with
\begin{align}
	x&\equiv a^2q_{\rm max}^2\approx 2255\times Z^{-2/3}\left(\frac{m_{\chi}}{100\text{MeV}}\right)^2\left(\frac{v}{10^{-3}}\right)^2.\nonumber
\end{align}
The common pre-factor corresponds to isotropic scattering, and the case specific factors capture how the different form factors alter the scattering kinematics. The distribution functions are presented in Fig.~\ref{fig:pdfcosa}. It is interesting to note the contrasting behaviors of contact interactions and long range forces. For contact interactions, the atomic form factor, or in other words the screening of the nuclear charge on large scales, determines the scattering kinematics. Backwards scattering is generally favored, especially for masses below 100 MeV. For larger masses, the screening becomes less relevant, and we obtain isotropic scattering, as expected. In contrast, long range forces favor forward scattering due to the cross section being proportional to $q^{-4}$, especially for masses above 100 MeV. Below that the charge screening renders the distribution more and more isotropic. The heavier (lighter) and faster (slower) a DM particle is, the more isotropic the scattering becomes for contact interactions (long range forces). The kinematics of electric dipole interactions show a somewhat intermediate behavior, favoring backwards scattering for light DM particles, where the charge screening dominates the scattering kinematics, and forward scattering for heavier DM, caused by the DM form factor being proportional to $q^{-1}$.

The procedure to find the critical cross section, above which a detection experiment does not rule out parameter space is unchanged to previous works~\cite{Emken:2018run}. Starting at the conventional bound on the cross section for a given DM mass, we systematically increase the DM-electron scattering cross section, where we use Eq.~\eqref{eq:sigma ratio} to obtain the nuclear cross section $\bar{\sigma}_p$, and simulate the nuclear stopping and reflection by the overburden. The sample of velocities at detector depth is then used to estimate the underground DM~speed distribution function via Kernel Density Estimation\footnote{Kernel Density Estimation~(KDE) is a non-parametric method to estimate the underlying~PDF of a data set. For more details, we refer to Appendix~A.1 of~\cite{Emken:2018run}.}, which in turn can be used to compute the attenuated signal rates and constraints. The cross section needs to be increased with decreasing step sizes when approaching the critical cross section in order not to overshoot beyond the critical value. Near the critical cross section, the probability of a DM particle reaching the detector becomes extremely low, and the signal rate at the detector decreases very steeply. This is typically the regime where also the simulation time increases rapidly.

\subsection{Electronic Stopping Power and Analytic Method}
In this section, we consider the contributions by the electrons of a medium to the overall DM stopping power of Eq.~\eqref{eq: stopping power total}. The DM particles can interact with the electrons either elastically (in which case the entire atom recoils), which determines $S_a$ in Eq.~\eqref{eq: stopping power total}, or through inelastic processes like atomic ionization or excitations in crystals, which determines $S_e$ in Eq.~\eqref{eq: stopping power total}.  For the inelastic processes, it is useful to recall the basic kinematic properties when the DM scatters off an electron: the \textit{typical} recoil energy of an outer-shell bound electron in an atom or the high-level valence electrons in a crystal is a few eV~\cite{Essig:2015cda}.  It is therefore much easier to excite an electron in a crystal, whose band gap is $\mathcal{O}$(eV), compared to ionizing an electron in an atom, whose ionization energy is $\mathcal{O}$(10~eV).  

We first consider the case of Earth's crust as the medium, before considering the atmosphere. The Earth's crust mostly consists of molecules that have ionization energies of $\sim \mathcal{O}(10 \hspace{1mm} \text{eV})$. The only crystal contained in Earth's crust at an appreciable amount ($\sim 6.7 \%$ of the crust by mass)~\cite{rudnick2003composition} is Iron(II) oxide (FeO), which has a band gap of $\sim 2.5 \hspace{1mm} \text{eV}$. The band structure and wave functions of electrons in FeO have not yet been calculated. However, the band gap of FeO is similar to silicon's~($\sim 1.1$~eV) for which the excitation form factors are available~\cite{Essig:2015cda}. To estimate the electronic stopping power of the crust, we model it therefore for simplicity as silicon and take the number density of the silicon atoms to be equal to the number density of FeO in the Earth's crust. The electronic stopping power due to inelastic scatterings off electrons in the Earth's crust can be estimated as
\begin{equation}\label{eq:ionizationstopping}
S_e(\mathbf{x},E_\chi)=\frac{n_{\rm cell} \bar{\sigma}_{e} \alpha m_{e}^{2}}{\mu_{\chi e}^{2}v_{\chi}v_{\rm rel}} \int \dd E_{e} E_{e} \int \frac{\dd q}{q^2} F_{\rm DM}(q)^2 \left| f_{\rm crystal} (q,E_e) \right|^2 \Theta[v_{\chi}-v_{\rm min}(E_{e},q)]\, ,
\end{equation} 
as derived in Appendix~\ref{app:derivation}. 

In addition to inelastic processes, the DM particles may also lose energy through elastic scatterings on electrons. In this case we assume that  DM scatters elastically on a bound electron such that the \textit{entire} atom recoils. The electronic DM stopping power contribution by elastic scatterings off atoms can be estimated via   
\begin{equation}\label{atomicstopping}
S_a(\mathbf{x},E_\chi)=\sum_i  n_{i}Z_{i}\int_{0}^{q_{max}^2} \dd q^2\, \left(\frac{q^{2}}{2m_{i}}\right) \frac{\dd \sigma_{e}}{\dd q^2}\,,
\end{equation} 
where $n_i$, $Z_i$, and $m_i$ are the number density, atomic number, and mass of the $i^{\text{th}}$ target, respectively. 

The two surface experiments (SENSEI and SuperCDMS) for which we will calculate limits consisted of a silicon target.  For these two experiments, we need to consider DM scattering in the Earth's atmosphere. The atmosphere of the Earth consists mostly of the gases $N_2$ ($\sim 76\%$ by weight) and $O_2$ ($\sim 23\%$ by weight). The ionization energies of these diatomic molecules are $\sim \mathcal{O}(10 \hspace{1mm} \text{eV})$. 
However, since the thresholds of the semiconductor target experiments are of $\mathcal{O}$(1~eV), the critical cross section is essentially determined solely by $S_a$ given in Eq.~\eqref{atomicstopping}, since only elastic scattering off electrons in atoms is capable of slowing down the incoming~DM from $\mathcal{O}$(10~eV) to below~$\mathcal{O}$(1~eV). 

In summary, for an experiment placed underground, the electronic stopping power is the sum of both inelastic and elastic interactions with electrons. For the atmosphere, we neglect inelastic processes for experiments with semiconductor targets and consider only elastic DM-electron scatterings. 

\paragraph{Analytic method}
In order to estimate the critical cross section above which DM particles that interact exclusively with electrons are stopped from producing observable signals in direct-detection experiments, we use an analytic approach. 
Let $f_{\rm halo}(v)$ be the initial DM velocity distribution and $f(v)$  be the final DM velocity distribution at the detector placed at some depth $d$ below the atmosphere or Earth's crust. Consider a dark matter particle with initial velocity $v_{0}$, which gets slowed down to a final velocity of $v_{d}$ at the detector. We have 
\begin{eqnarray}\label{am3} 
 \int_{E_0}^{E_d}\dd E_\chi\, \left(\frac{\dd \langle E_\chi\rangle}{\dd x}\right)^{-1} &=& d,
\end{eqnarray} 
where $E_{0}$ = $\frac{1}{2}m_{\chi} v_{0}^{2}$ is the DM particle initial kinetic energy, $E_{d}$ = $\frac{1}{2}m_{\chi} v_{d}^{2}$ is its final energy, and $\frac{\dd \langle E_\chi\rangle}{\dd x}$ is the stopping power. For $S_e$ (the stopping power from inelastic scattering) given in Eq.~(\ref{eq:ionizationstopping}), we simply find the critical cross section for which the fastest moving DM particle ($v_0 \sim$784 km/s) slows down sufficiently so that its kinetic energy at the detector is below the detector's threshold energy, i.e.,~$E_d$=$E_{\rm{thr}}$. This kind of energy/speed cutoff criterion has been applied in several studies, see e.g.~\cite{Starkman:1990nj,Kouvaris:2014lpa,Emken:2017erx}.

For $S_a$ (the atomic scattering stopping power) given in Eq.~(\ref{atomicstopping}), we use a slightly more detailed approach (for $S_e$, the 
following approach turned out to be difficult to evaluate numerically).  We find the final velocity distribution by using the conservation of the particle flux, 
\begin{eqnarray}\label{am4} 
f(v_{d})\, v_{d} \dd v_{d} &=& f_{\rm halo}(v_{0}) \, v_{0} \dd v_{0}\,.
\end{eqnarray} 
The DM speed distribution enters the direct detection rates through the $\eta$ function defined in~\eqref{eta}. We use this velocity distribution at the position of the detector to calculate the number of events expected in the direct detection experiments as a function of cross section using Eq.~\eqref{semiconductor}. The critical cross section is then found by increasing the cross section to a point where the stopping effect brings down the expected number of events to the required value at 95\% confidence level.

For an experiment underground, we individually compute the limits based on the ionization and atomic scattering stopping power. Then we choose the stronger limit of these two (i.e., the limit with the lower cross section for the upper boundary). For the semiconductor experiments on the surface, we calculate the limit based on atomic scattering alone.

\section{Results}
\label{s:results}

\begin{figure*}
	\centering
	\includegraphics[width=0.5\textwidth]{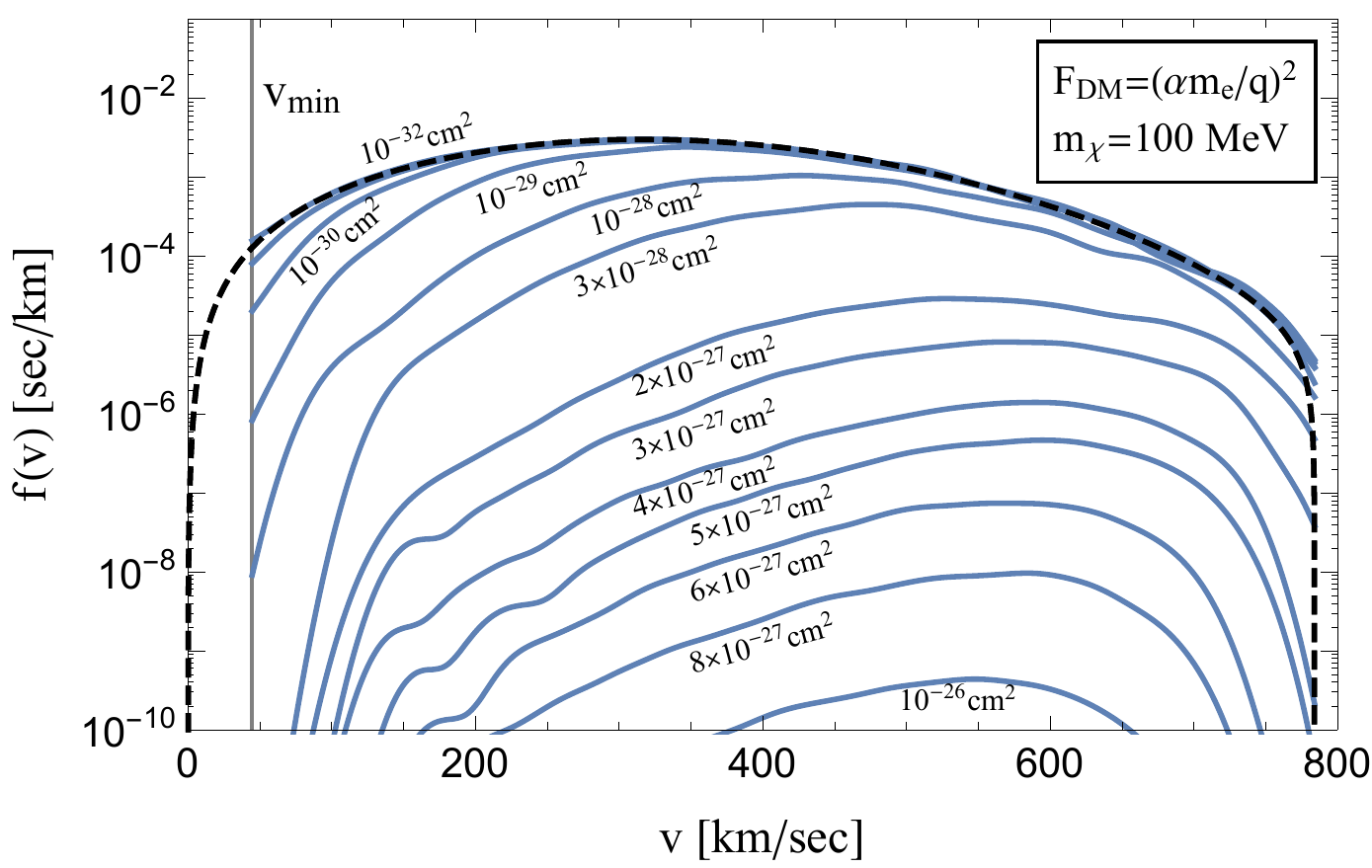}
	\includegraphics[width=0.48\textwidth]{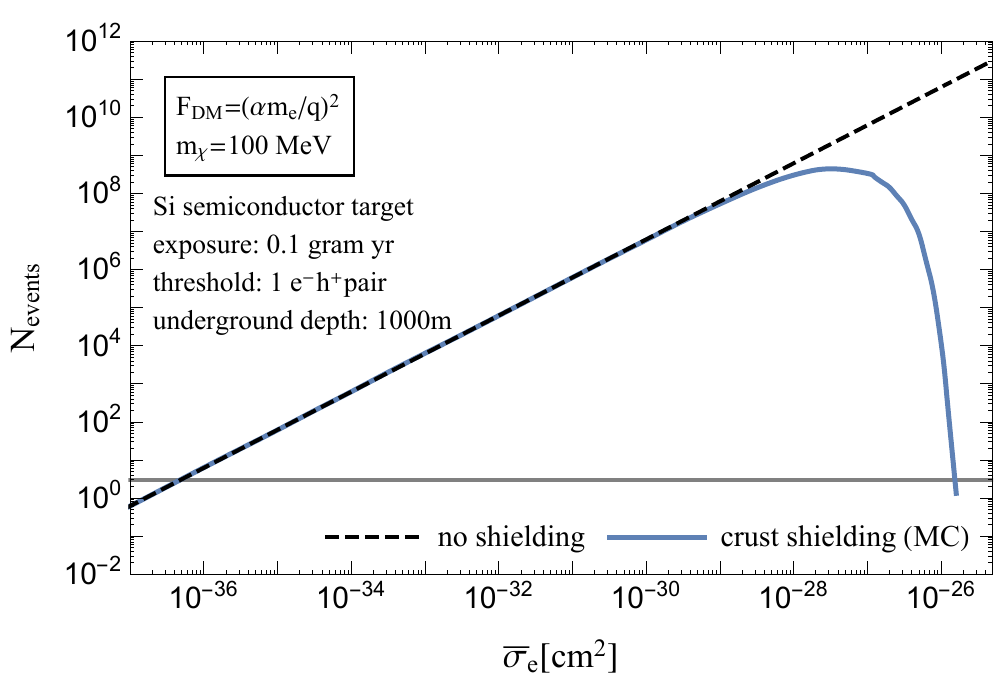}
\caption{The left panel shows the distortion of the DM~speed distribution due to underground scatterings on nuclei in 1~km of rock overburden obtained with MC~simulations. The labels indicate the respective values of~$\overline{\sigma}_e$. The right panel shows the resulting attenuation of the number of expected events in a generic semiconductor experiment. The grey line indicates $\sim$3 events, i.e. the number of signals corresponding to the 95\% CL exclusion bound in the absence of background events. }
\label{fig: underground pdfs and events}
\end{figure*}

The effect on the local DM~distribution of scatterings on terrestrial nuclei in the Earth crust is shown in the left plot of Fig.~\ref{fig: underground pdfs and events}. Here, we assume a DM~mass of 100~MeV, an ultralight mediator, and an underground depth of 1~km. It demonstrates the depletion of the number of observable DM~particles, i.e. particles with speed above~$v_{\rm min}$. We can also see that the population of slow particles gets depleted more efficiently, which can be attributed to the DM~form factor suppressing large momentum transfers. These distortions of the DM~distributions have an impact on the expected signals at underground detectors, as shown in the right plot of Fig.~\ref{fig: underground pdfs and events}. Above a certain cross section, the total number of signal starts to drop quickly, as the Earth crust turns essentially opaque to DM~particles\footnote{In~\cite{Mahdawi:2018euy}, it was argued that the planar modelling of the shielding layers of the overburden is not applicable for atmospheric shielding and leads to an underestimation for the number of detectable particles by up to a factor of 2. While this might be accurate, the corresponding effect on the critical cross section is negligible, which can be seen in the right panel of Fig.~\ref{fig: underground pdfs and events}.}. 

\begin{figure*}
 \centering 
\includegraphics[width=0.49\textwidth]{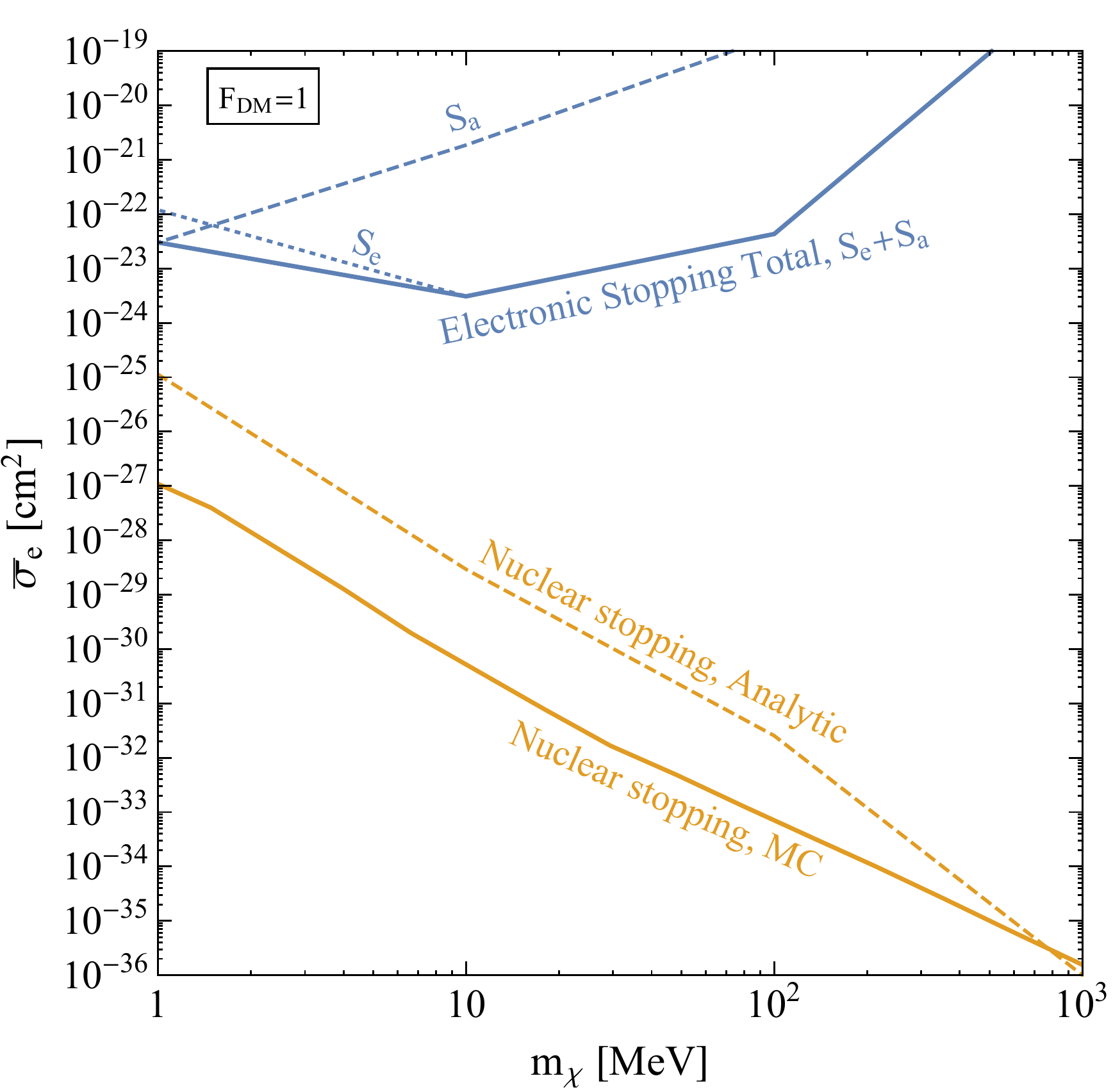}
~\includegraphics[width=0.49\textwidth]{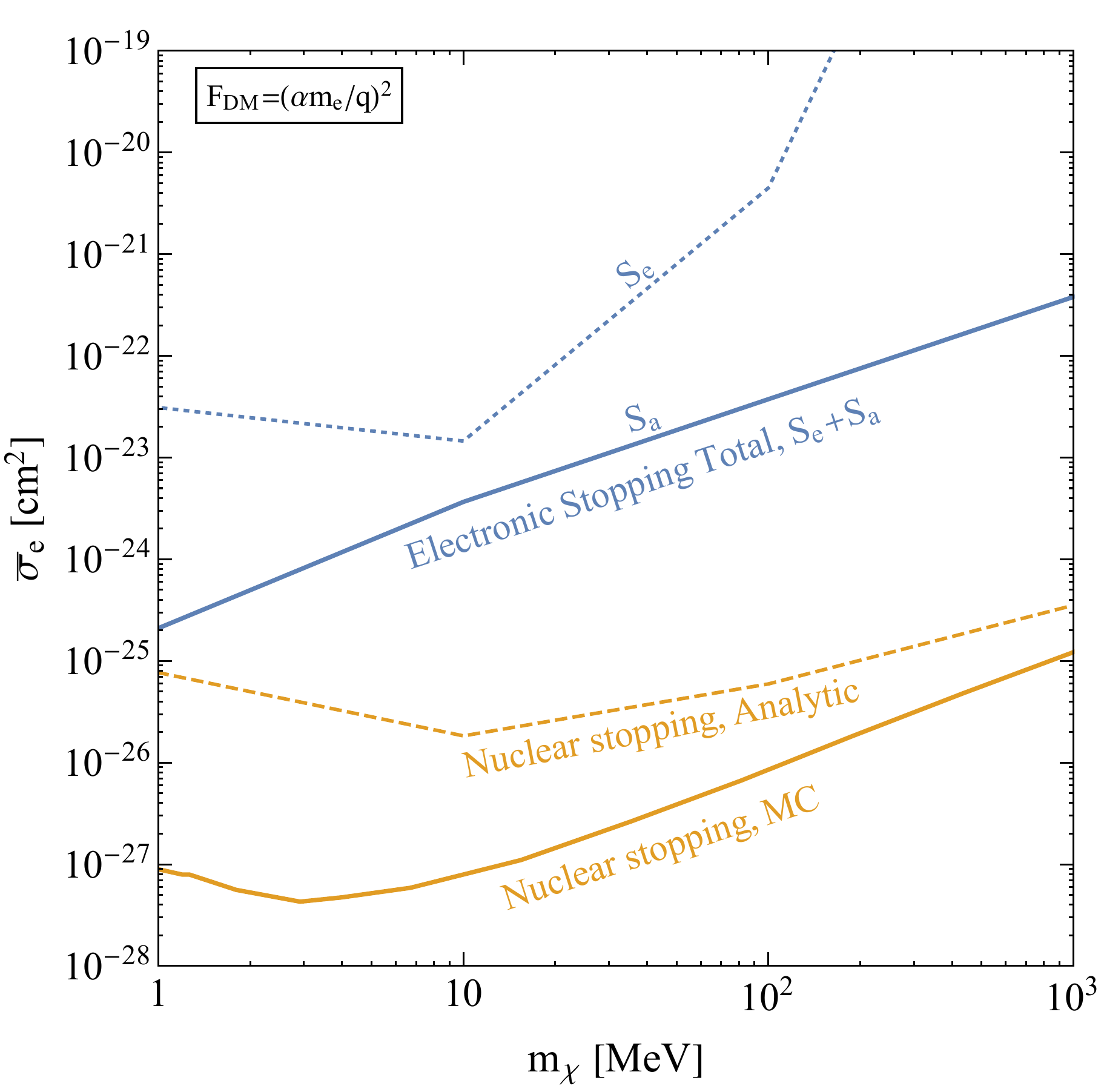}
\caption{Comparison of dark matter stopping through interactions with electrons (blue curves) versus interactions with nuclei (yellow curves) for $F_{\text{DM}}=1$ (left panel) and $F_{\text{DM}}$ = $(\alpha m_e / q)^2$ (right panel). We silicon as the target, a threshold of one electron-hole pair (i.e. an energy threshold of 1.1.~eV), an exposure of 100 gram-year, and that the detector is placed 2km underground, the depth of SNOLAB. The critical cross sections based on the ionization stopping power (dotted line, labelled $\text{S}_{\text{e}}$) and atomic scattering (dashed line, labelled $\text{S}_{\text{a}}$) are shown separately, while the critical cross section based on the total stopping power is shown by the thick line. We also show a comparison between the critical cross section due to nuclear interactions obtained with an analytic (dashed line) and a MC method (thick line).  }
\label{fig:estopping}
\end{figure*}

In Fig.~\ref{fig:estopping}, we compare the shielding effects due to the different stopping processes by showing the corresponding upper boundaries of the exclusion limits of a generic DM-electron scattering experiment with silicon semiconductor target with a threshold of one electron and an exposure of 100 gram-year. We assume the experiment to be at SNOLAB ($\sim$2000 m underground). Across the whole mass range of interest, we find that the critical cross section due to elastic scatterings on nuclei falls significantly below the corresponding value set by electronic interactions. We conclude that elastic scatterings on nuclei dominate the Earth shielding. It is therefore legitimate in this case to determine the critical cross section neglecting the DM-electron interactions.

Furthermore, Fig.~\ref{fig:estopping} contains a comparison between the critical cross section due to nuclear interactions obtained with analytic and MC methods. For low DM masses, the analytic approach over-estimates the critical cross section, as it fails to take scattering deflections into account. As the mass approaches~$\sim$1 GeV, forward scattering dominates more and more as discussed in the previous section. Deflections play a minor role, and the assumptions of the stopping equation, specifically a continuous energy loss along a straight path, are more accurate for heavier DM particles. This is reflected by the fact that the two values for the critical cross sections converge around 1 GeV. For larger masses the analytic approach should yield accurate results.  This is fortunate, since the MC simulations become computationally very expensive for large masses, as the number of scatterings of particles that reach the detector increases significantly.

Finally, based on previous results for contact interactions~\cite{Mahdawi:2017cxz,Emken:2018run}, one might presume that the critical cross section obtained by using the analytic stopping power yields a \textit{conservative} result. Fig.~\ref{fig:estopping} clearly shows that this is not the case in general. In the case of sub-GeV DM, the analytic approach clearly fails for low masses, overestimating the critical cross section significantly and falsely suggesting that parts of the parameter space are excluded, while MC~simulations reveal them to be perfectly viable.

\subsection{Constraints}

\begin{figure}[!t]
\vspace{4mm}
\includegraphics[width=0.5\textwidth]{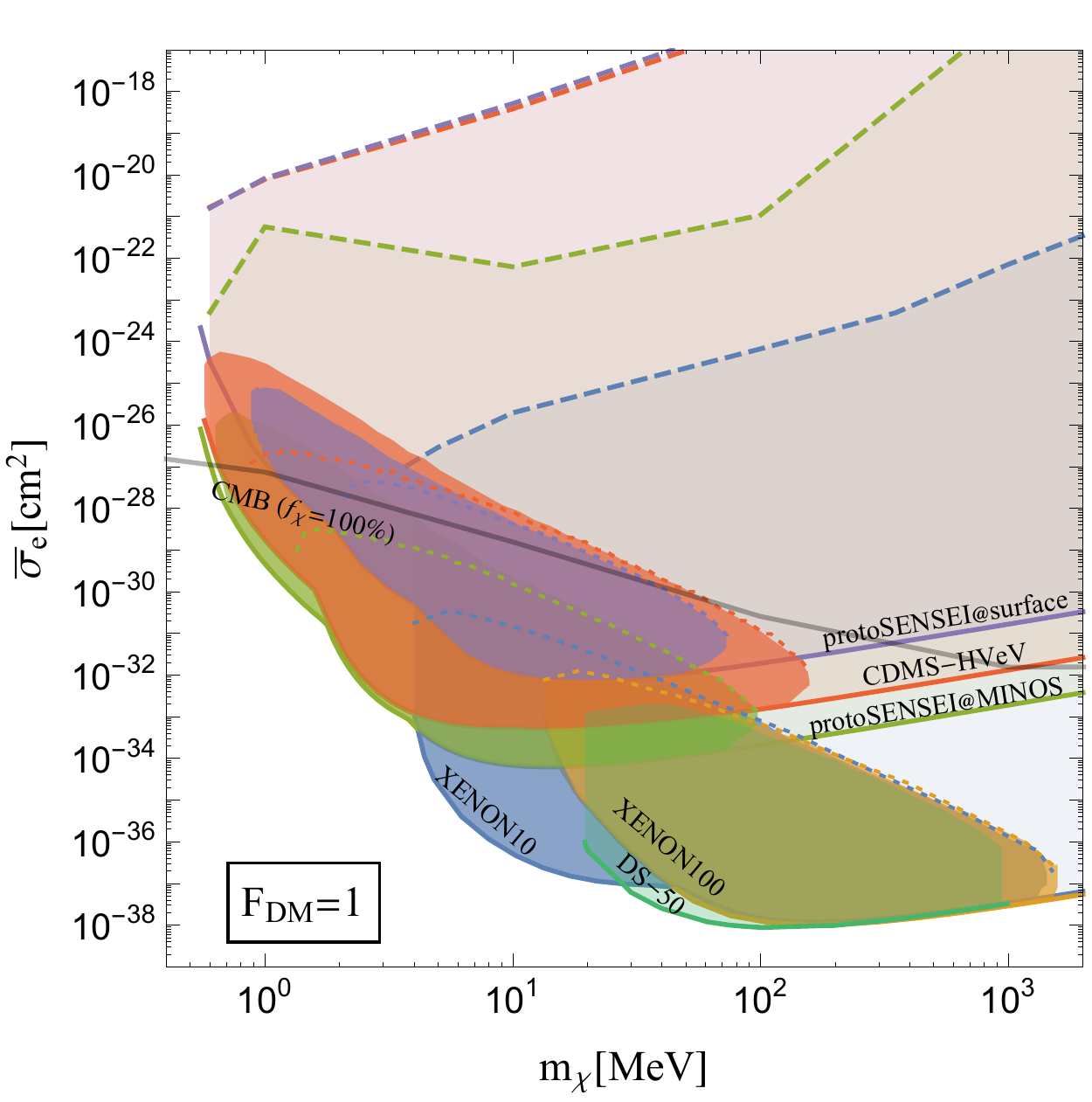}
~\includegraphics[width=0.5\textwidth]{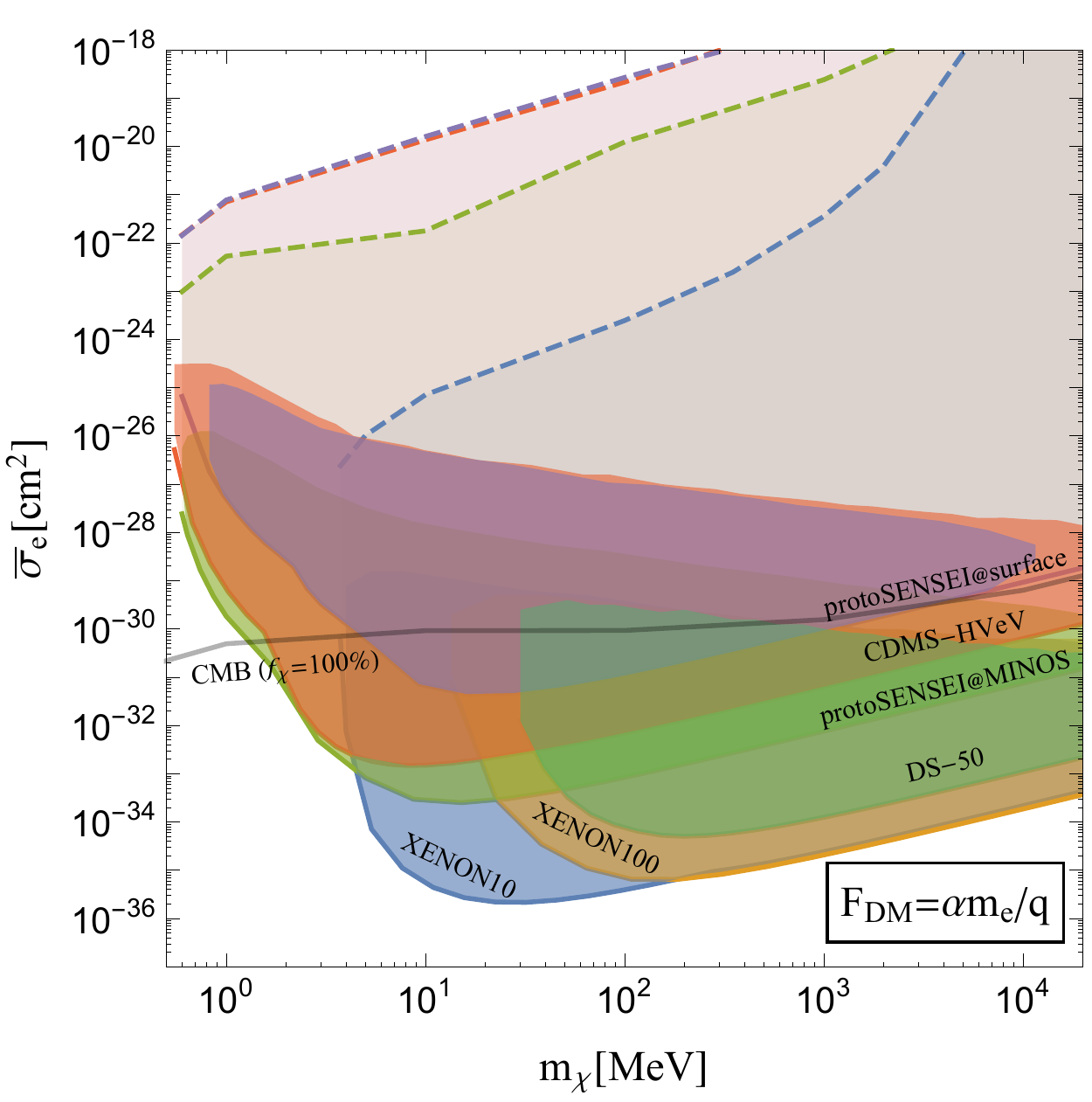}
\vspace{-3mm}
\\
\begin{minipage}[t]{0.5\textwidth}
\mbox{}\\[-\baselineskip]
\includegraphics[width=\textwidth]{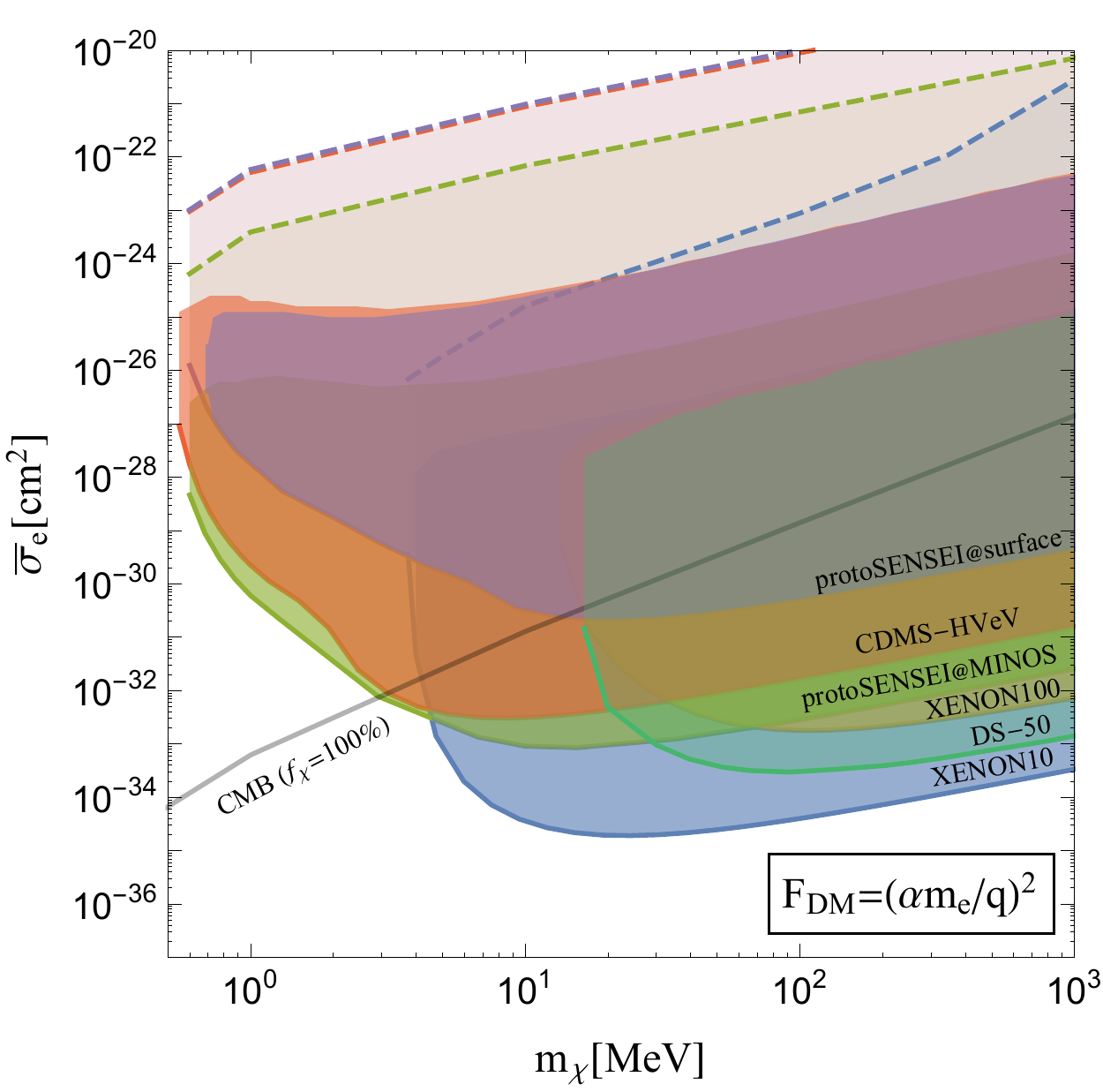}
\end{minipage}
\hfill
\begin{minipage}[t]{0.45\textwidth}
\mbox{}\\[-\baselineskip]
\vspace{-15.5pt}
\caption{Constraints on the DM-electron scattering cross section for XENON10~\cite{Angle:2011th,Essig:2017kqs}, XENON100~\cite{Aprile:2016wwo,Essig:2017kqs}, SENSEI (protoSENSEI@surface~\cite{Crisler:2018gci} and protoSENSEI@MINOS~\cite{Abramoff:2019dfb}), DarkSide-50~\cite{Agnes:2018oej}, and CDMS-HVeV (as  for SENSEI, we show the constraint without Fano-factor fluctuations, which is the upper boundary of the constraint band shown in~\cite{Agnese:2018col}). The dashed lines show the upper boundary for the constraints obtained with electronic stopping only. The additional dotted lines in the upper left panel show the constraints for contact interactions in the absence of charge screening. The gray solid lines show the strongest CMB constraints for $f_{\chi}$=$100\%$~\cite{Xu:2018efh}.}
\label{fig:constraints}
\end{minipage}
\end{figure}
 
In Fig.~\ref{fig:constraints}, we present the currently leading direct-detection constraints on sub-GeV DM from experiments sensitive to DM-electron scattering, including XENON10~\cite{Angle:2011th,Essig:2017kqs}, XENON100~\cite{Aprile:2016wwo,Essig:2017kqs}, SENSEI (protoSENSEI@surface~\cite{Crisler:2018gci} and protoSENSEI@MINOS~\cite{Abramoff:2019dfb}), SuperCDMS~(``CDMS-HVeV'')~\cite{Agnese:2018col}, and DarkSide-50~\cite{Agnes:2018oej}. The details of each of these experiments are summarized in Appendix~\ref{app:experiments}. The dark-shaded regions show the constraints on sub-GeV DM models with contact and long-range interactions mediated by a dark photon (left top and left bottom plot, respectively), and an electric dipole moment interaction (right plot).  Constraints on DM with electron-only couplings are shown in light-shaded regions (enclosed by dashed lines).  It comes as no surprise that the experiments that took data on the surface (the two SENSEI runs as well as CDMS-HVeV) exclude much larger cross sections than the experiments that took data underground (XENON10, XENON100, and DarkSide-50).  

In Fig.~\ref{fig:constraints}, we also show the constraints coming from baryon interactions with the CMB for each of the three models~\cite{Xu:2018efh} (for related work, see e.g.~\cite{Chen:2002yh,Dvorkin:2013cea,Gluscevic:2017ywp,McDermott:2010pa,Dolgov:2013una,Dubovsky:2003yn,Slatyer:2015jla,Ali-Haimoud:2015pwa,Boddy:2018wzy}).  These limits assume that the interacting DM component, $\chi$, makes up the entire DM abundance 
($f_{\chi} \equiv \Omega_\chi/\Omega_{\rm DM} = 100\%$).  While converting the DM-baryon cross section limits derived in~\cite{Xu:2018efh} to the DM-electron cross section, we have used the Debye screening length of the baryon plasma to regulate the divergence for small momentum transfers in the cases of long range interaction mediated by a dark photon and an electric dipole moment interaction. 

\begin{figure*}
\centering
\begin{minipage}[t]{0.57\textwidth}
\mbox{}\\[-\baselineskip]
\includegraphics[width=\textwidth]{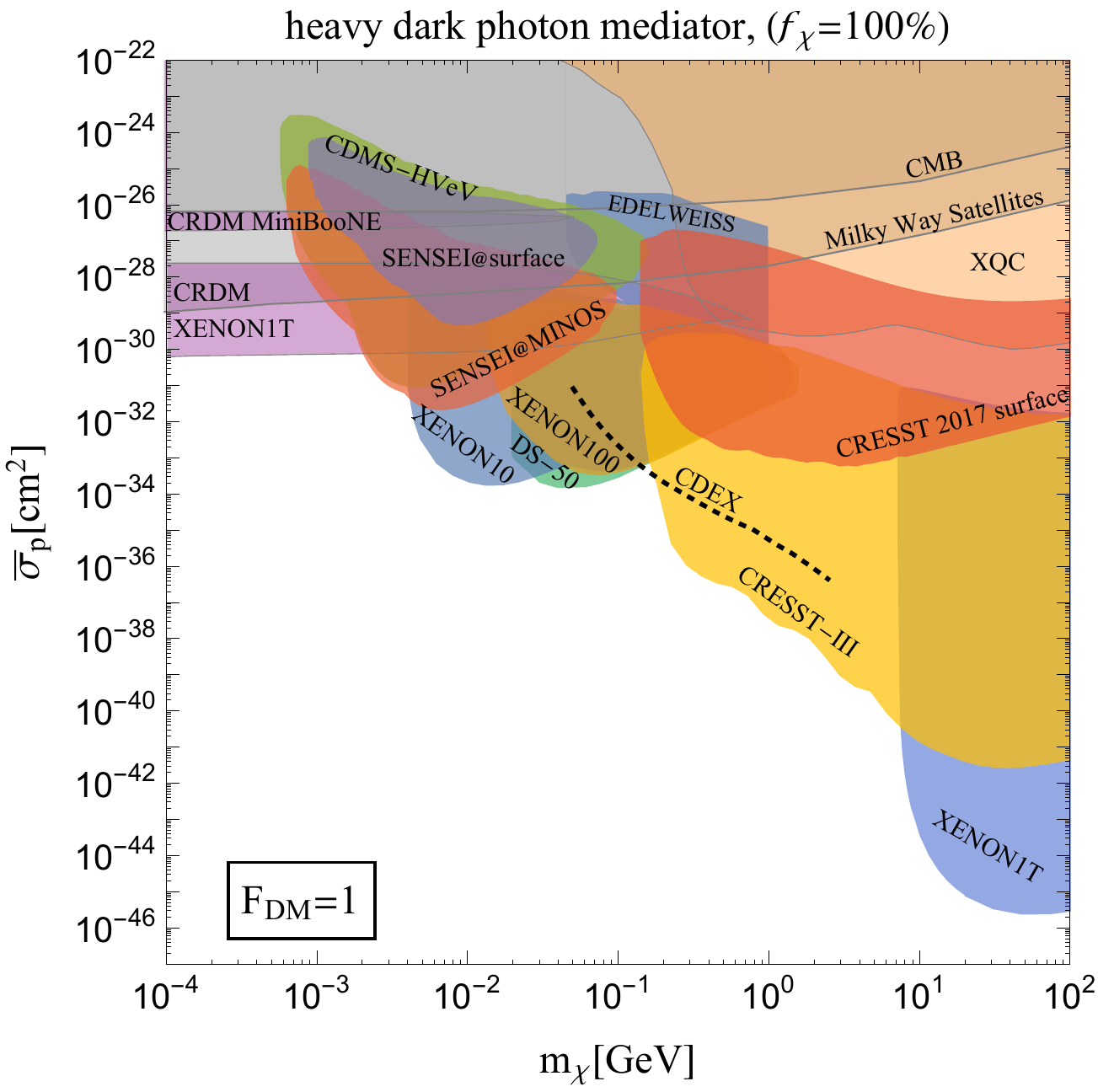}
\end{minipage}
\hfill
\begin{minipage}[t]{0.42\textwidth}
\mbox{}\\[-\baselineskip]
\vspace{-15.5pt}
\caption{Direct-detection, cosmological, and astrophysical constraints on the DM-proton cross section for contact interactions mediated by a dark photon. In addition to the bounds derived in this paper (see Fig.~\ref{fig:constraints}, top left), we also show constraints from nuclear recoil DM-searches by XENON1T~\cite{Aprile:2017iyp}, CRESST-III~\cite{Abdelhameed:2019hmk,Abdelhameed:2019mac}, the CRESST 2017 surface run~\cite{Angloher:2017sxg} as obtained in~\cite{Emken:2018run}, and Migdal effect based bounds from EDELWEISS~\cite{Armengaud:2019kfj} and CDEX~\cite{Liu:2019kzq}, together with constraints from the X-ray Quantum Calorimeter experiment (XQC)~\cite{Mahdawi:2018euy}, cosmic-rays (``CRDM'')~\cite{Bringmann:2018cvk}, CMB~\cite{Xu:2018efh}, and Milky-Way satellites~\cite{Nadler:2019zrb}.  We do not show collider or beam-dump bounds, but see e.g.~\cite{Batell:2009di,Essig:2013vha,Batell:2014mga}.  
}
	\label{fig:constraints sigma p}
\end{minipage}
\end{figure*}

We see that no parameter space remains open between the CMB limits and the direct-detection constraints, except for a small region for a DM particle interacting with a heavy mediator for $m_{\chi} \gtrsim 100~\text{MeV}$.  However, for these higher masses, and for a heavy dark-photon mediator, other constraints will close this parameter space, see Fig.~\ref{fig:constraints sigma p}.  
In Fig.~\ref{fig:constraints sigma p}, we have translated all the bounds shown in Fig.~\ref{fig:constraints} from $\overline\sigma_e$ to the DM-proton scattering cross section, $\overline \sigma_p$, using Eq.~(\ref{eq:sigma ratio}).  We also show bounds from nuclear-recoil search experiments (scaled by a factor of 4 to account for the dark photon interacting only with protons) and astrophysical searches.   

\subsection{Projections}

\begin{figure}[!t]
\vspace{4mm}
\includegraphics[width=0.5\textwidth]{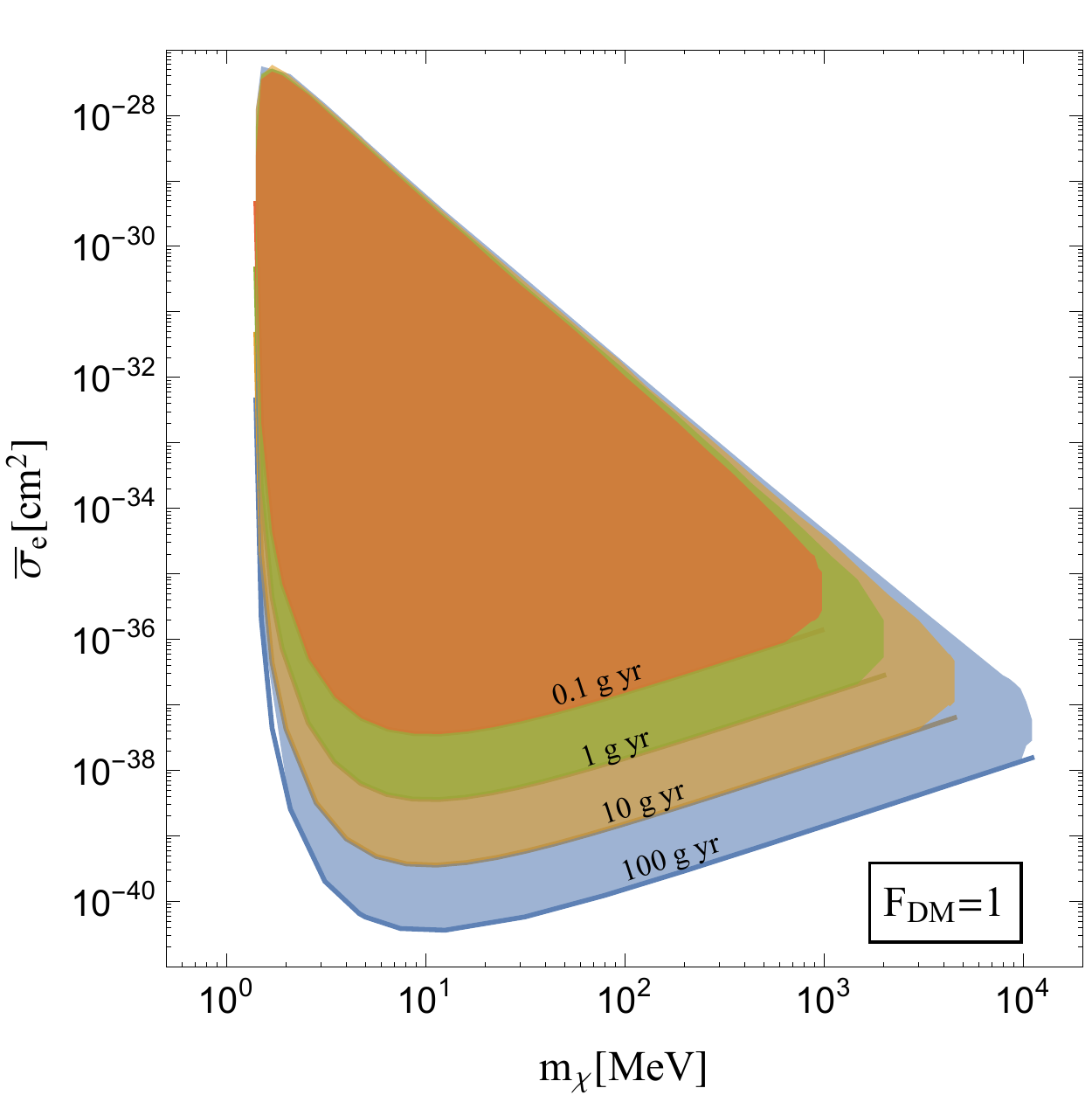}
~\includegraphics[width=0.5\textwidth]{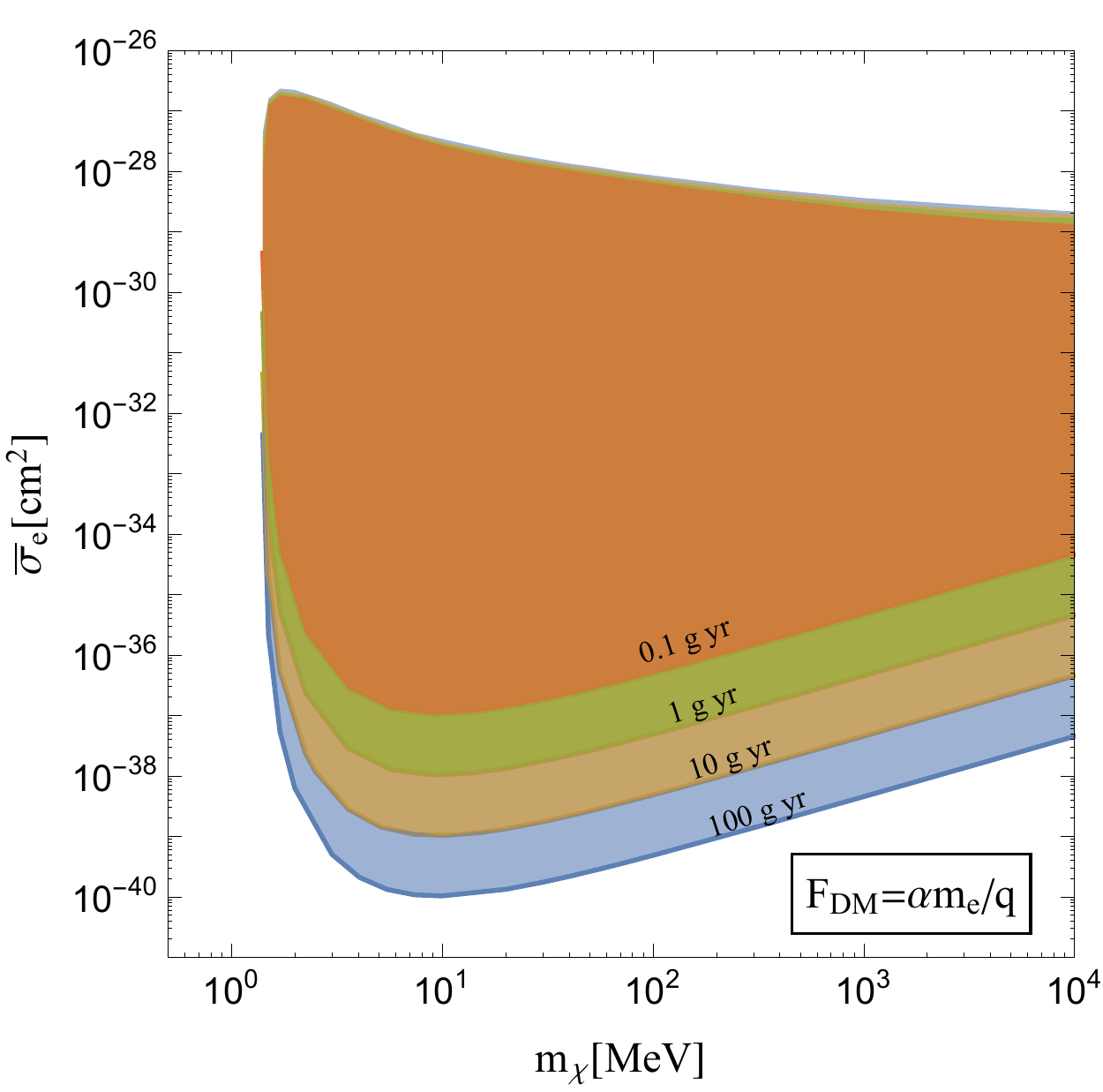}
\vspace{-3mm}
\\
\begin{minipage}[t]{0.5\textwidth}
\mbox{}\\[-\baselineskip]
\includegraphics[width=\textwidth]{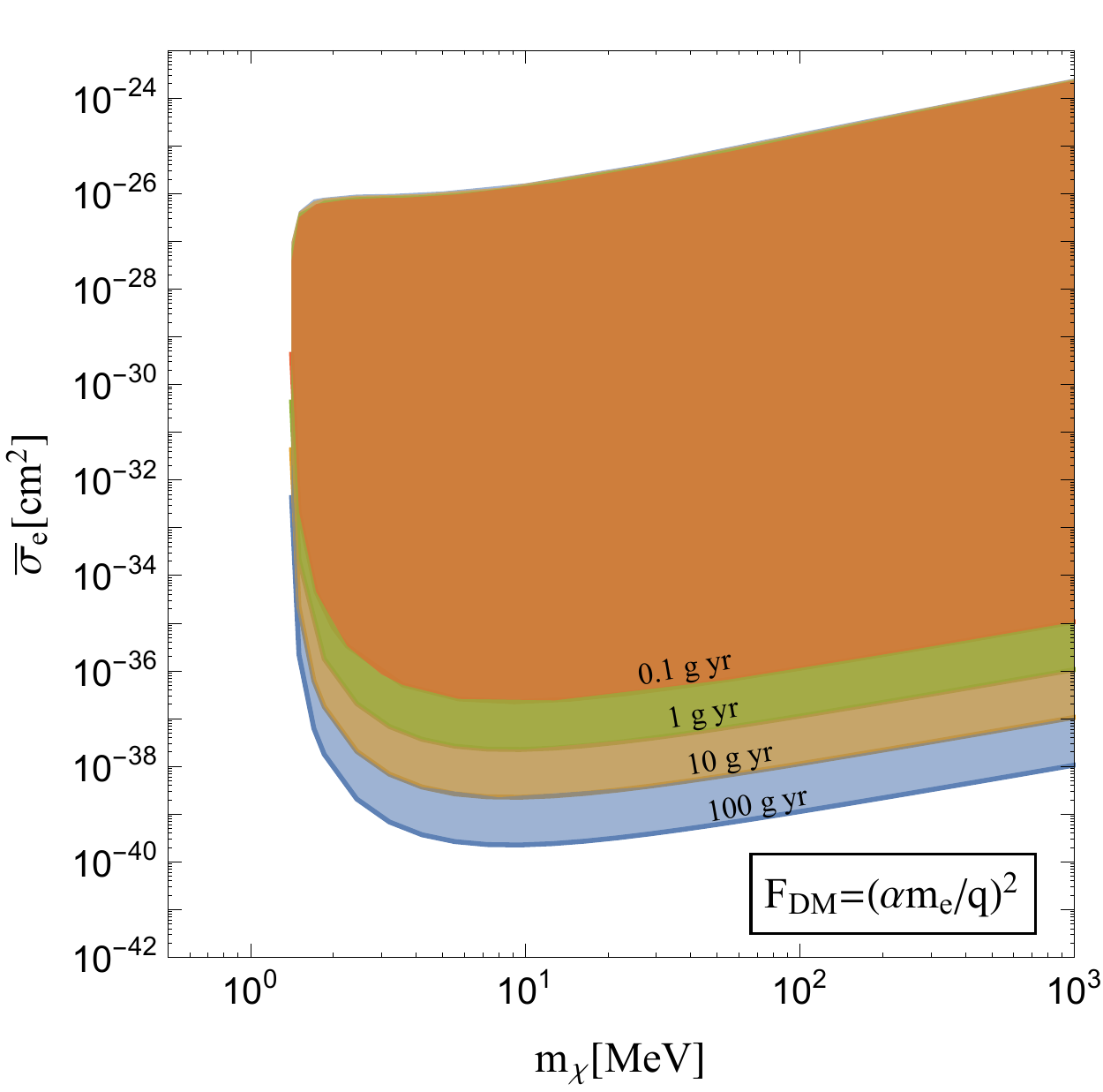}
\end{minipage}
\hfill
\begin{minipage}[t]{0.45\textwidth}
\mbox{}\\[-\baselineskip]
\vspace{-15.5pt}
\caption{Projected sensitivity at 95\%~CL on the DM-electron scattering cross section versus DM mass for various exposures.  We assume the detector consists of a silicon target, zero background events, a threshold of two electron-hole pairs, and an underground depth of 100~m.}
\label{fig:Projection_exposure}
\end{minipage}
\end{figure}

\subsubsection{Scaling of Upper Boundary of the Critical Cross Section with Exposure}\label{subsubsec:exposure}  
A direct-detection experiment excludes a band of cross sections. Probing lower cross sections, and extending the exclusion band towards weaker interactions can be achieved by increasing the exposure, i.e., by running experiments with larger targets for longer times (assuming backgrounds do not scale with exposure).  We here investigate how much the constraints at the upper boundary improve from having a larger exposure. In~Fig.~\ref{fig:Projection_exposure}, we present projected constraints at 95\% CL for a generic underground semiconductor detector with a silicon target. We fixed the threshold to~2 electron hole pairs, the underground depth to~100~m, and assume that no background events have been observed. Furthermore, we vary the exposure between~0.1 and 100~gram-years to study how the upper boundary scales exactly with exposure. We find, of course, that the lower boundary decreases linearly with exposure. However, the upper boundary is insensitive to the exposure. An increase of four orders of magnitude in exposure increases the critical cross section by less than~$\sim$60\%. Above a certain interaction strength, the expected number of triggered events in a detector falls extremely fast, and the overburden becomes effectively opaque to~DM. The properties of the experiment play only a minor role. In the case of contact and electric dipole interactions, there is no sensitivity above some DM mass for a given exposure, but a higher exposure extends the sensitivity to higher masses. 

\begin{figure}[!t]
\vspace{4mm}
\includegraphics[width=0.5\textwidth]{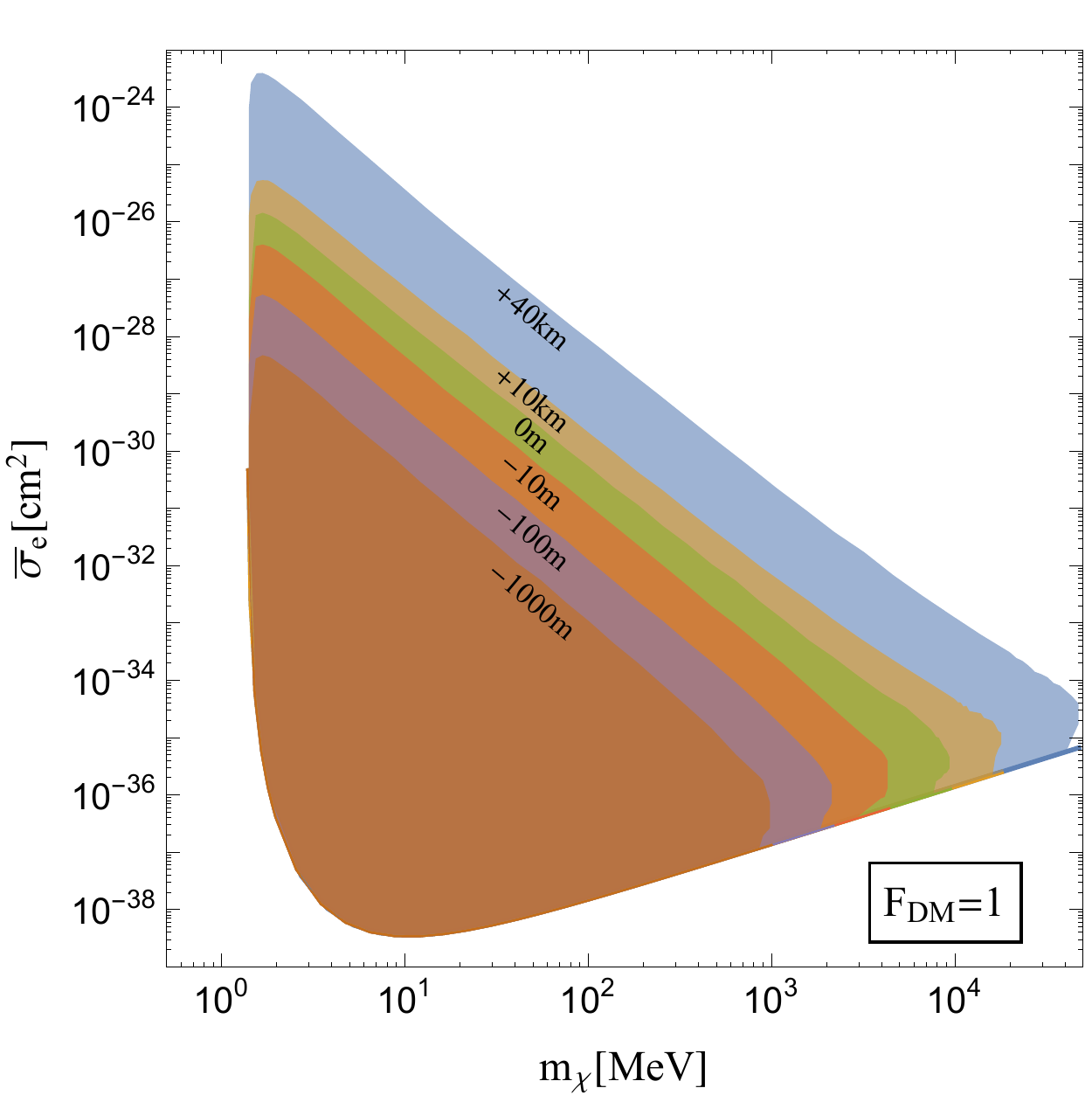}
~\includegraphics[width=0.5\textwidth]{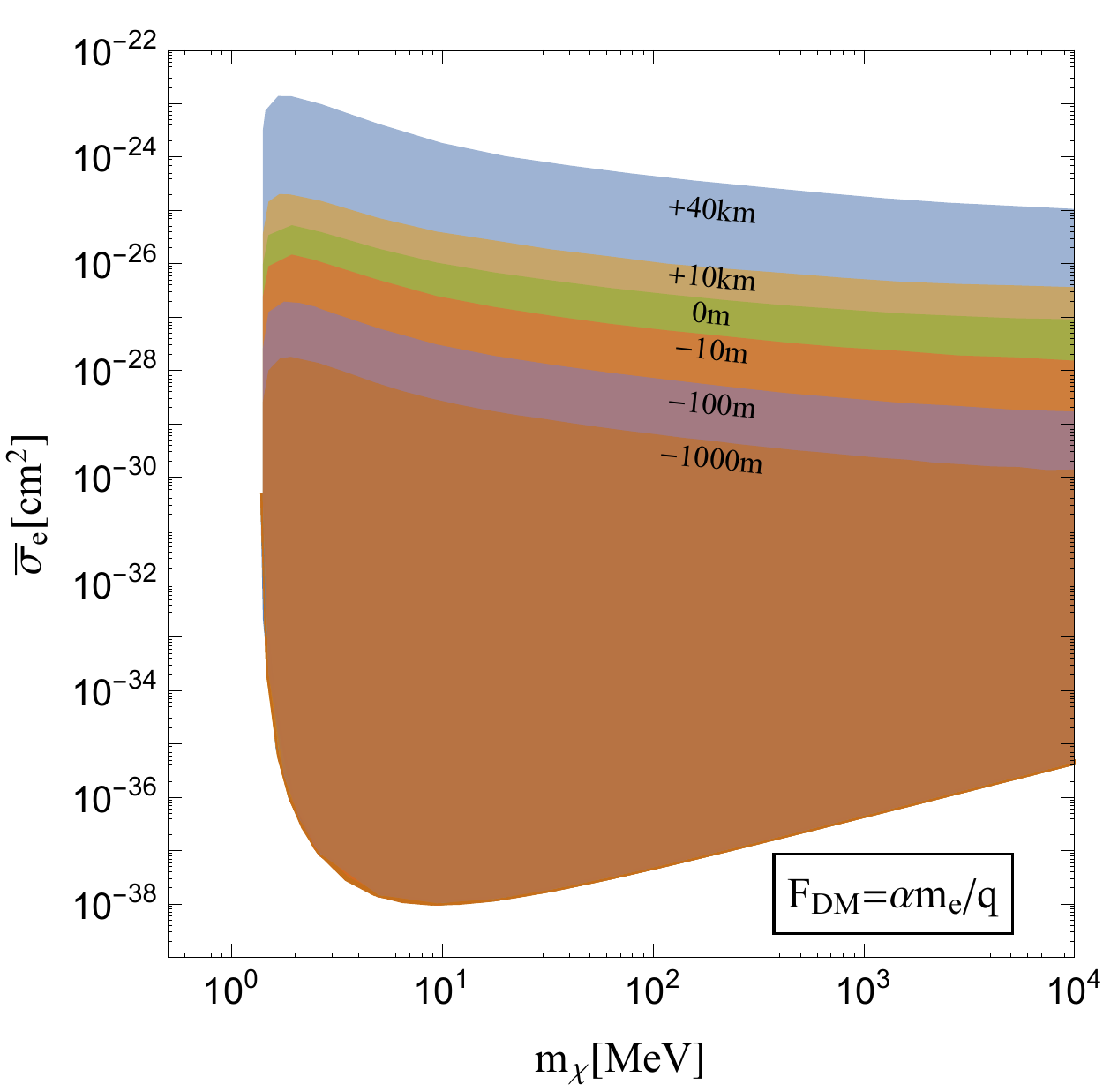}
\vspace{-3mm}
\\
\begin{minipage}[t]{0.5\textwidth}
\mbox{}\\[-\baselineskip]
\includegraphics[width=\textwidth]{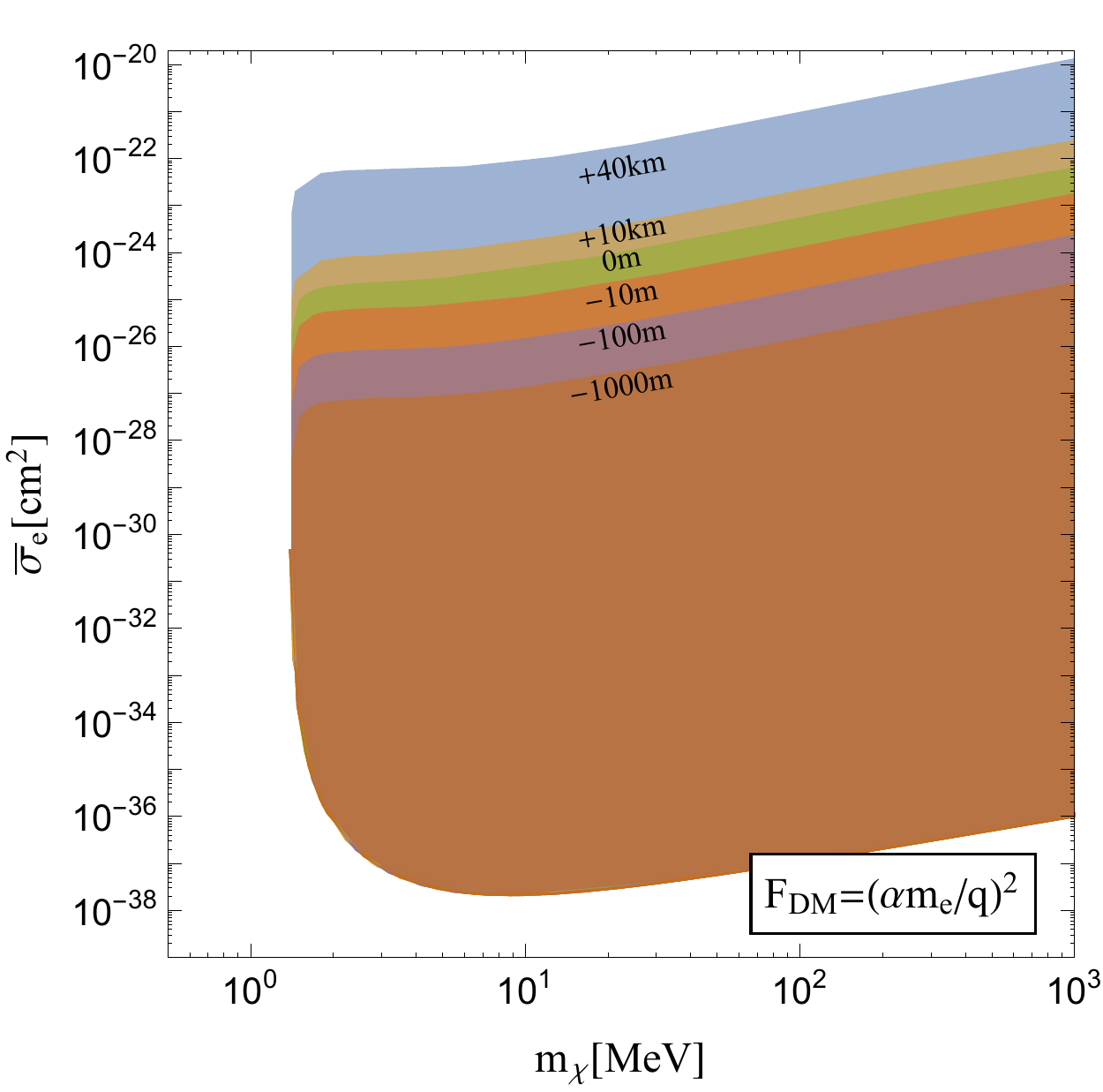}
\end{minipage}
\hfill
\begin{minipage}[t]{0.45\textwidth}
\mbox{}\\[-\baselineskip]
\vspace{-15.5pt}
\caption{Projected sensitivity at 95\%~CL on the DM-electron scattering cross section versus DM mass for experiments located at several underground depths and above-ground altitudes. We assume a silicon target, zero background events, a detector threshold of two electron-hole pairs, and an exposure of 1 gram-year.}
\label{fig:Projection_depth}
\end{minipage}
\end{figure}

\subsubsection{Scaling of Upper Boundary of the Critical Cross Section with Detector Depth} 
The second question is how much the critical cross section can be increased by moving the experiment to a shallower site. In Fig.~\ref{fig:Projection_depth}, we present the exclusion bands' scaling in terms of the underground depth, or overground altitude. Again, we assume a silicon detector with a threshold of 2 electron hole pairs, and a fixed exposure of 1~gram-year without background events. We vary the underground depth between 1000~m underground up to 40~km altitude, which could be reached by using a balloon experiment. For the underground depths of 1000~m and 100~m we only take the Earth crust into account. For the remaining depth and heights, we also include the atmosphere. The atmosphere at 40~km altitude has a very low density, and has a comparable stopping power to packaging material that may surround the detector. Here, we assume to have additional layers of steel (2~mm) and copper (1~mm). 

We find, as a rule of thumb, that for each order of magnitude decrease in underground depth, we gain one order of magnitude of sensitivity towards DM with large interaction strengths. By moving the experiment to an elevated location, the critical cross section can be pushed up further by up to one more order. Furthermore, it could be highly advantageous to set up a balloon experiment or an experiment on a satellite, as the low density of the upper atmosphere or the absence of any atmospheric shielding, respectively, increases the experimental sensitivity to larger cross sections.  We will discuss this possibility further below.  

\subsubsection{Projections for SENSEI and DAMIC-M}
\begin{figure}[!t]
\vspace{4mm}
\includegraphics[width=0.5\textwidth]{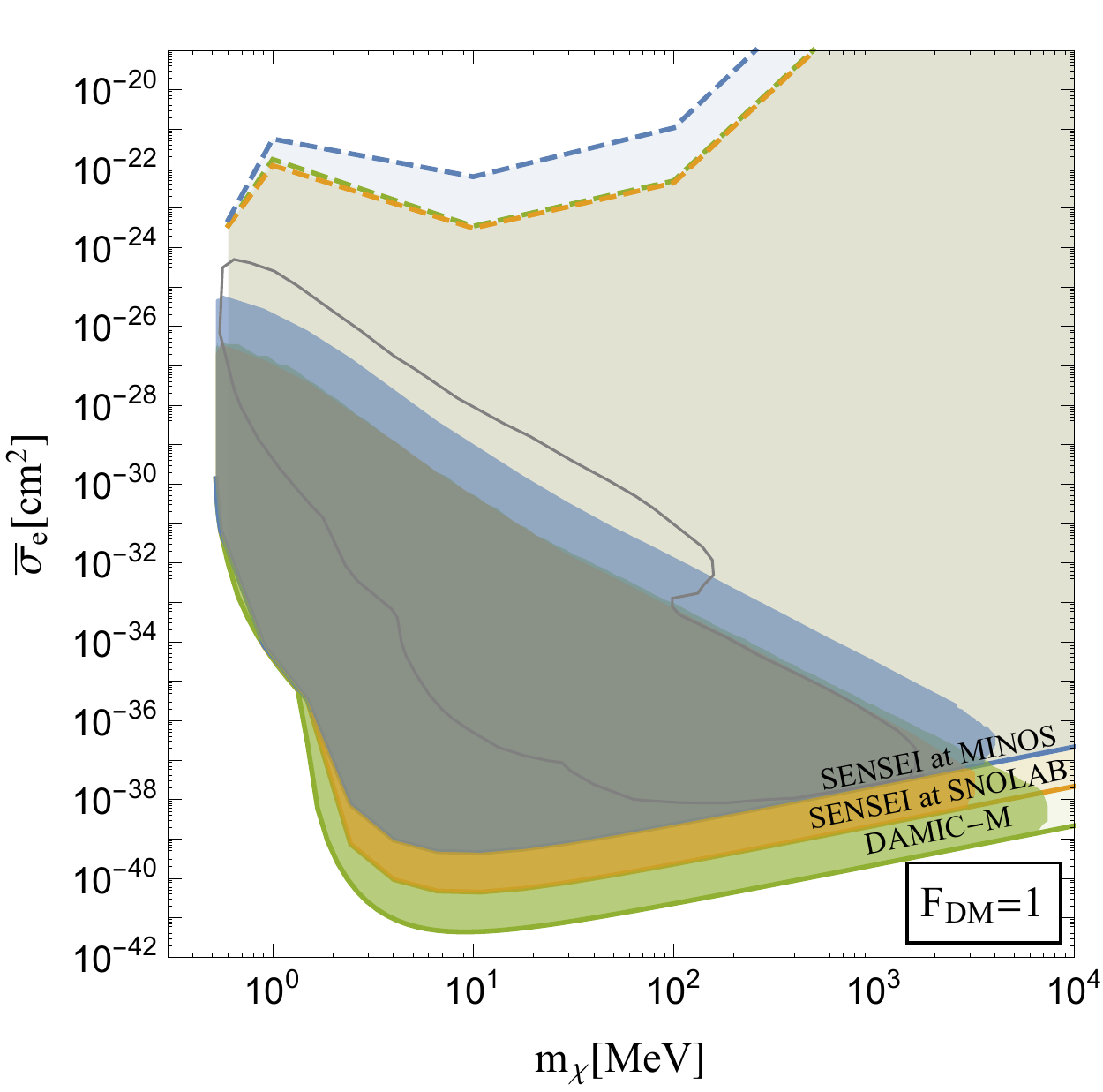}
~\includegraphics[width=0.5\textwidth]{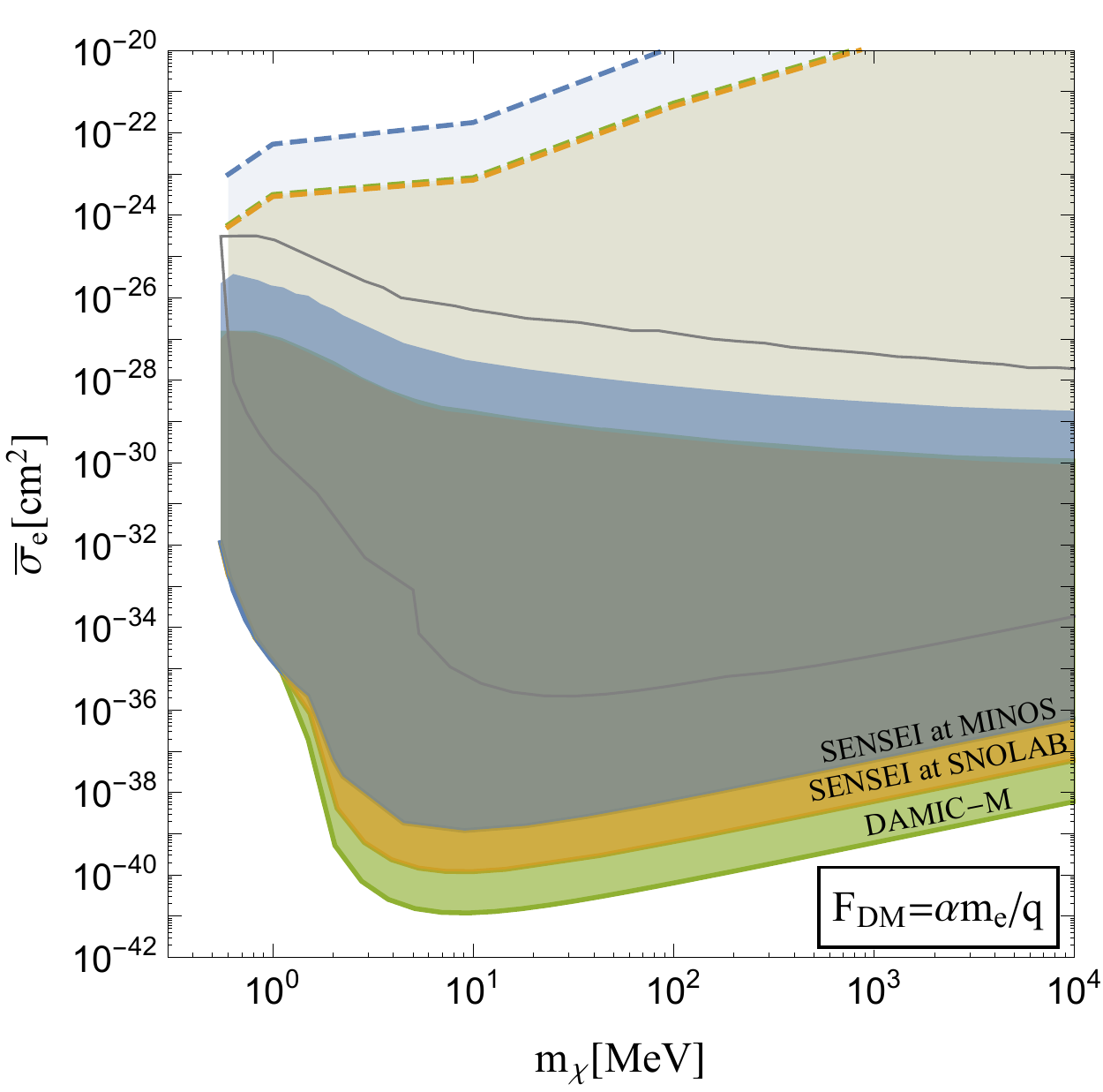}
\vspace{-3mm}
\\
\begin{minipage}[t]{0.5\textwidth}
\mbox{}\\[-\baselineskip]
\includegraphics[width=\textwidth]{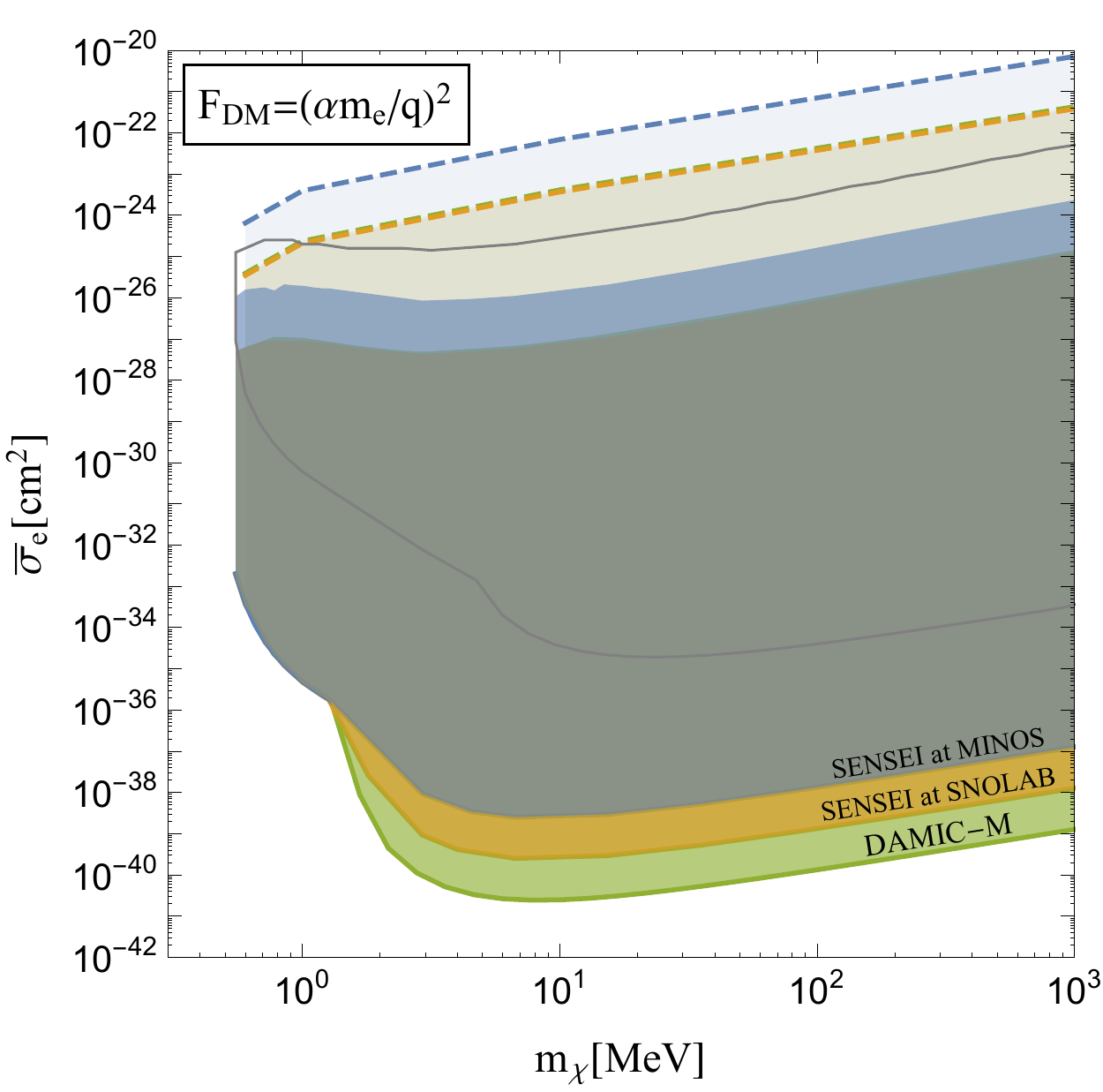}
\end{minipage}
\hfill
\begin{minipage}[t]{0.45\textwidth}
\mbox{}\\[-\baselineskip]
\vspace{-15.5pt}
\caption{Projected sensitivities at 95\% CL on the DM-electron scattering cross section for SENSEI at MINOS~(107~m underground), and SNOLAB~(2000~m underground), and DAMIC-M in Modane~(1780~m underground). The dashed lines show the upper boundary for the projections obtained with electronic stopping only. The existing constraints from Fig.~\ref{fig:constraints} are included as a grey outline.}
\label{fig:ProjectionSenseiDamic}
\end{minipage}
\end{figure}

We show projections in Fig.~\ref{fig:ProjectionSenseiDamic} for the upcoming DM-electron scattering experiment SENSEI and the longer-term DAMIC-M, which both use a silicon semiconductor target, and determine their sensitivity to sub-GeV DM with large interaction strengths. For SENSEI at MINOS, SENSEI at SNOLAB, and DAMIC at Modane, we assume exposures of 10, 100, and 1000~gram-year, respectively, together with a threshold of one electron hole pair. For the background, we assume an observed number of events in the first electron bin of $10^3$, $10^4$, and~$10^5$ events, respectively. This is an optimistic estimate of the number of expected ``dark current'' events~\cite{Tiffenberg:2017aac}. While DAMIC-M will probe weaker DM-electron interactions due to its large exposure (assuming control over backgrounds), a SENSEI run at the shallower MINOS is more sensitive to large interactions. It will, however, not be able to compete at large cross sections with the SENSEI surface run using a prototype Skipper CCD and the CDMS-HVeV surface run.  

\subsubsection{Balloon and satellite experiments} 
For large cross sections, where underground and surface experiments are not sensitive to DM due to terrestrial effects of the crust and atmosphere, it could be beneficial to run a direct-detection experiment at high altitude or on a satellite. We entertain the idea of a small semiconductor detector placed on a balloon or a satellite.  

A balloon experiment could probe higher cross sections than a surface experiment due to the low density of the upper atmosphere, while a detector on a satellite would have no atmospheric shielding to contend with.  In both cases, the detector would need to be surrounded by some minimal amount of packaging material in order to regulate the detector temperature and (in the case of the balloon) provide a vacuum.  However, one side of the detector could be shielded with very little material, while the other sides would need to house the power supply, electronics, and other detector components, which combined would provide a significant amount of shielding.  

For both a balloon- or satellite-borne experiment, we expect a large modulation of the signal rate. 
For example, one could point the detector side containing the least amount of shielding towards Cygnus, which is the direction of motion of the Sun.  
The expected DM signal rate would then depend strongly on the position of the detector  relative to the Earth, since the Earth would block a large fraction of the DM flux coming from Cygnus.  A characteristic signature of DM with large interaction strengths would be a strong modulation due to this shadowing effect~\cite{Collar:1992qc,Kouvaris:2014lpa}. The modulation phase, amplitude, and even frequency depend on the experiment's location. For a balloon-borne experiment we would expect a frequency of one per sidereal day, while a satellite-borne experiment could look for orbital modulations of higher frequencies. 
This modulation enables the potential not only to constrain DM with large interaction strengths, but also to \textit{discover} it.  
It would be a powerful discriminant against possible systematics as well as to modulations in the background rate induced by the rotating Earth or by the detector orbiting the Earth.  
In addition to the signal rate modulation caused by the motion of the Earth or the satellite around the Earth, one could also imagine rotating the detector away or towards the direction of Cygnus, which would allow for a controlled modulated signal rate and adds another powerful discriminant. 
This would be possible provided that the experimental apparatus itself causes a significant attenuation of the DM~flux. 
Then the exact time-dependent orientation of the experiment with respect to the galactic frame has to be taken into account.
Of course, in the event of a signal, having both a balloon-borne and satellite-borne detector could help with determining the DM properties.

The signal modulation can be estimated without the need of MC simulations. 
Instead, we assume that the DM~particles passing the Earth on their way to the detector get shielded completely, while the shielding of the experimental apparatus itself can be neglected. 
This is of course only accurate in a certain regime, but this regime includes the parameter space directly above the exclusions from terrestrial experiments.
Hence, we compute the local DM speed distribution at~$\mathbf{x}$ by simply filtering out those particles that would have passed through the Earth to reach the location,
\begin{align}
	f(\mathbf{x},v) &= \int \dd \Omega\; v^2 f(\mathbf{v}) \times \mathfrak{S}(\mathbf{x},\mathbf{v})\, ,
	\intertext{where the shadowing function is defined as}
	 \mathfrak{S}(\mathbf{x},\mathbf{v})&=\begin{cases}
		0\, ,\quad\text{if }\left|\mathbf{x}+\lambda \mathbf{v}\right|=r_\oplus \text{ has a solution }\lambda<0\, ,\\
		1\, ,\quad\text{otherwise.}
	\end{cases}
\end{align}
Therefore, we only include the stopping by the Earth for this estimate. 
We consider the following two concrete examples of high-altitude direct detection experiments with a silicon semiconductor target. 

\begin{enumerate}
	\item A geostationary balloon-borne detector launched in the northern hemisphere to  an altitude of 30~km over Pasadena, California~(34.1478${}^\circ$N,~118.1445${}^\circ$W).\footnote{The time dependent position vector of a location on Earth expressed in the galactic frame can be found in Eq.~(A.27) of~\cite{Emken:2017qmp}.  We also note that in practice a balloon may only be at high altitude for 
	$\mathcal{O}$(1~hour), and one could launch two balloons 12~hours apart to capture the maximum and minimum of the modulation amplitudes.} 
	\item A detector in orbit around the Earth, where we consider the International Space Station (ISS) as an example. The ISS orbits the planet at an altitude of~$\sim$400~km. The orbital, and therefore also the modulation period (assuming the detector's orientation does not change with respect to Cygnus) is about 90~minutes. Furthermore, having an orbital inclination of~$\sim$50${}^\circ$, the ISS is at times exposed to, and at other times shielded from, the~DM wind, leading to a large modulation amplitude.  
\end{enumerate}
Other options for both balloon- and satellite-borne detectors exist.  For example, CubeSats are another option, but would not qualitatively change the results presented here. In addition, other orbits for the satellite or other launch locations for a balloon may be more advantageous, as we discuss further below. 

\begin{figure*}
	\includegraphics[width=0.52\textwidth]{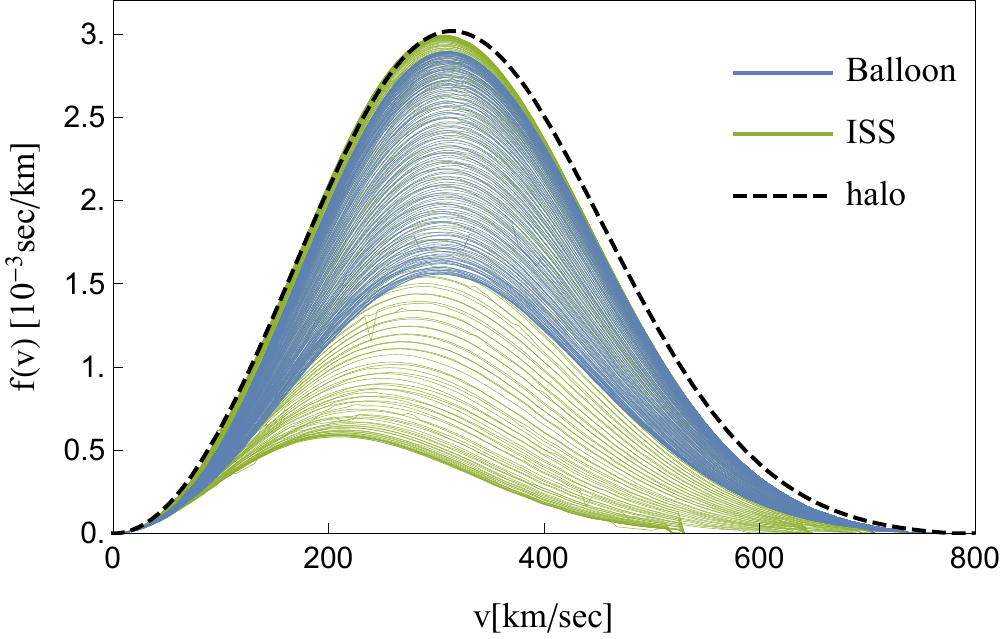}
	\includegraphics[width=0.48\textwidth]{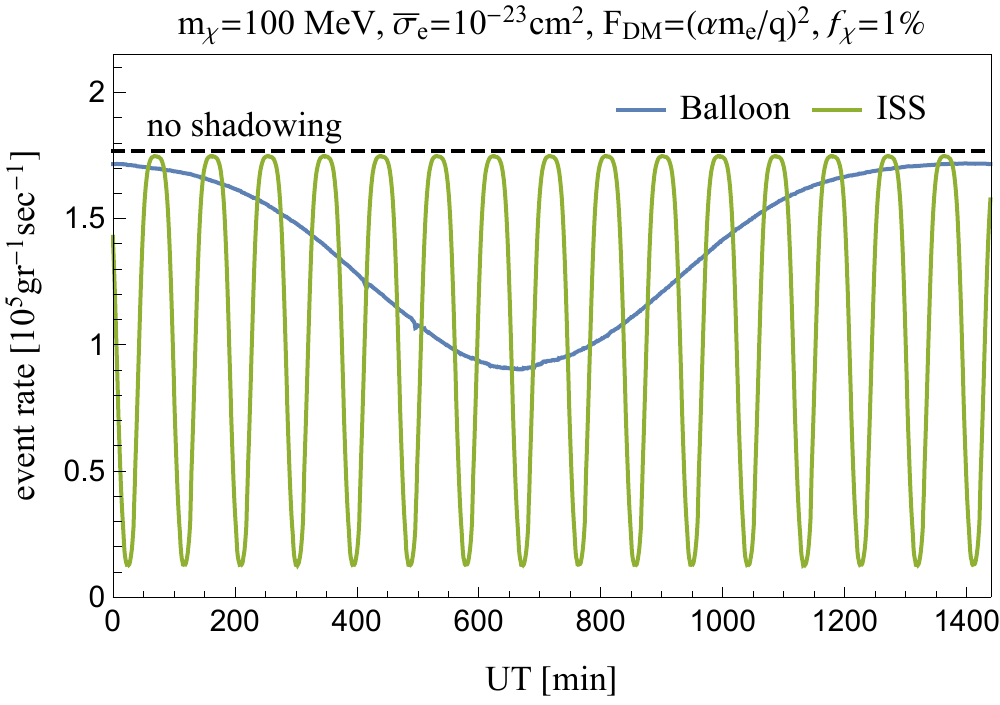}
	\caption{Orbital modulation of strongly interacting~DM at balloon and ISS borne semiconductor experiments during one day. The signal rate on the right corresponds to the parameters~$m_\chi=100$~MeV, and~$\overline{\sigma}_e=~10^{-23}~\mathrm{cm}^2$, an ultralight mediator, and a DM abundance of $f_\chi = 1\%$.}
	\label{fig: modulation}
\end{figure*}

The resulting DM speed distribution and event rates on a balloon and aboard the ISS are shown in Fig.~\ref{fig: modulation}. Here, we assume long-range interactions of a 100 MeV DM particle with a cross section of~$\overline{\sigma}_e=10^{-23}~\mathrm{cm}^2$ as an example, which lies just above the excluded region from the surface runs of SENSEI and CDMS-HVeV. The attenuated speed distributions show how the DM population gets depleted in the Earth's shadow. Especially the flux of fast DM~particles can be reduced to zero, as all particles from the high-speed tail approach the experiment from the same, particular direction. 

It is clear that the orbital signal modulation due to the Earth's shadow is significant and presents a powerful discovery signature. 
We define the fractional modulation as
\begin{align}
	f_{\rm mod}\equiv \frac{R_{\rm max}-R_{\rm min}}{R_{\rm min}+R_{\rm max}} = \frac{R_{\rm max}-R_{\rm min}}{2\langle R \rangle} \, ,
\end{align}
where~$R_{\rm min}$~($R_{\rm max}$) is the minimum~(maximum) rate along the orbit, and~$\langle R\rangle$ is the average signal rate. While the geostationary balloon-borne detector in the northern hemisphere would show a diurnal modulation with~$f_{\rm mod}\sim$30\%, a detector aboard a satellite or space station could expect high-frequency orbital modulations with~$f_{\rm mod}\sim$75\%. The fractional modulation is expected to increase for lower masses, as the detection of light~DM particles relies more on the high-speed tail of the distribution, which can be shielded off completely as discussed before.

These modulations could be the crucial feature to distinguish a signal from background, which in the absence of shielding layers is expected to be large. For a number of background events~$B$, and a given exposure~$\varepsilon$, we can estimate the~$5\sigma$ discovery reach of an orbital modulation due to a strongly interacting~DM particle of mass~$m_\chi$ by setting the signal-to-noise ratio with a flat background to
\begin{align}
	\frac{f_{\rm mod}S_{\rm tot}}{\sqrt{S_{\rm tot}+B}}=5\, ,
\end{align}
and solving the equation for the cross section~$\overline{\sigma}_e$~\cite{Essig:2015cda}.  Here,~$S_{\rm tot}\equiv \varepsilon \langle R \rangle$ is the total number of expected events. While the assumption of a flat background is optimistic, this simple formula is also rather conservative for a satellite-borne detector, as in practice one could search for the modulation amplitude using a large number of orbits.  In any case, this simple estimate will suffice for our purposes.  

\begin{figure*}
\centering
\begin{minipage}[t]{0.5\textwidth}
\mbox{}\\[-\baselineskip]
\includegraphics[width=\textwidth]{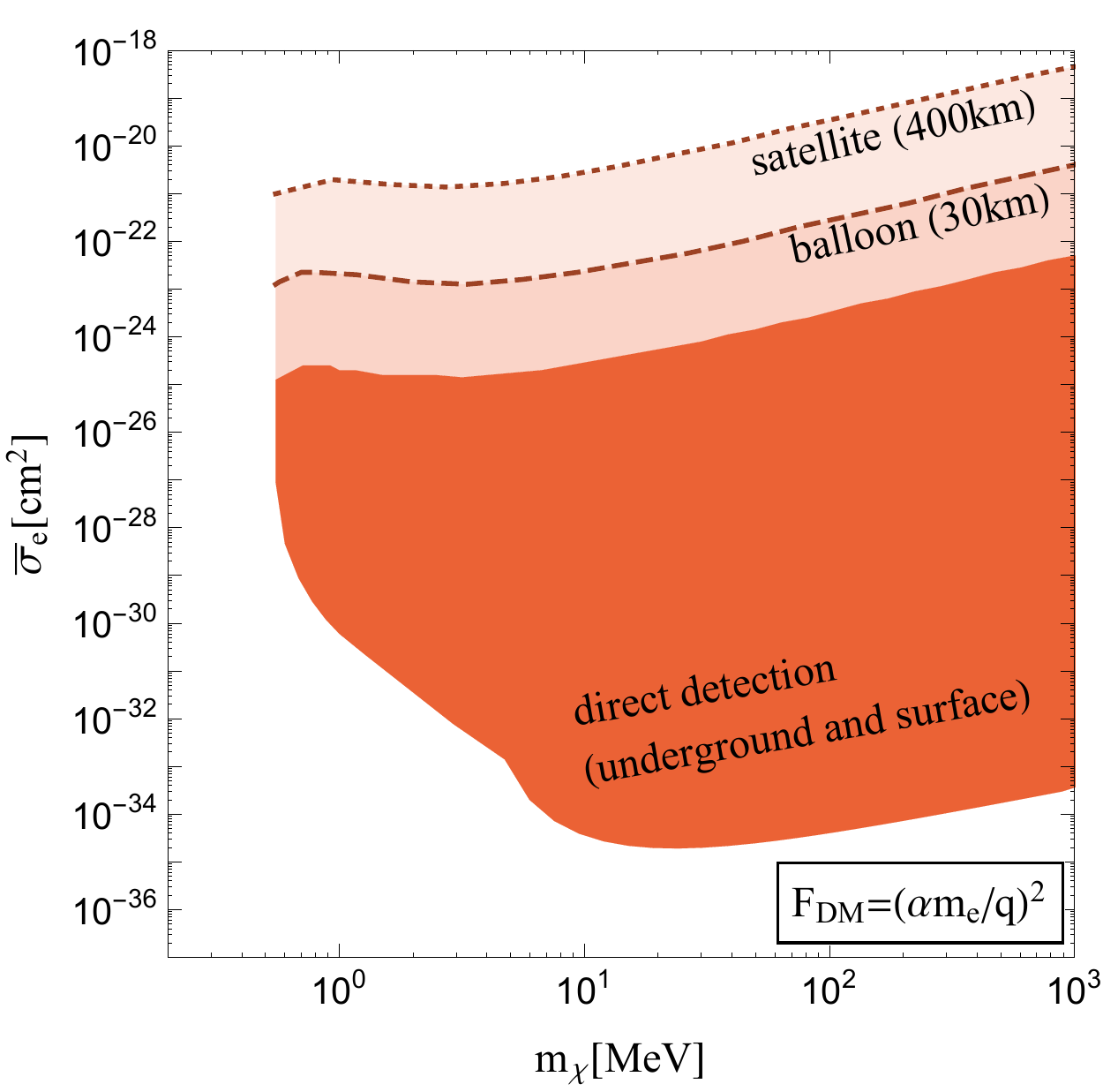}
\end{minipage}
\hfill
\begin{minipage}[t]{0.45\textwidth}
\mbox{}\\[-\baselineskip]
\vspace{-15.5pt}
\caption{ 
Discovery reach in dashed (dotted) lines for a silicon DM detector with single-electron sensitivity on a balloon (satellite) assuming an exposure of 1 gram-hour (0.1 gram-month) and $10^6$ ($10^9$) background events, assuming an ultralight dark-photon mediator.  The red region shows the direct-detection constraints derived in this paper from SENSEI, CDMS-HVeV, XENON10, XENON100, and DarkSide-50 (combined into one region).
}
	\label{fig:satellite-balloon}
\end{minipage}
\end{figure*}

For the balloon-borne experiment, we now assume an exposure of~1~gram-hour, and a background of~$10^6$ events. The corresponding parameters for the satellite experiment are taken to be 0.1~gram-months, and~$10^9$ background events.  These background numbers are chosen for purposes of illustration only.  Next, we compute the projected constraints and modulation discovery reach for the high-altitude experiments for the case of ultralight mediators. Both values are relatively insensitive to the background. We assume that the detector onboard the balloon is shielded by the upper atmospheric layers, as well as 5~mm of mylar, and 1~mm of copper.\footnote{The copper layer's density is set to~8.96 gram/$\text{cm}^3$, whereas mylar is modelled as a material with a density of 1.4~gram/$\text{cm}^3$, and composed of 62.5\% carbon, 33.3\% oxygen, and 4.2\% hydrogen~\cite{mylar}.} For the satellite-borne detector, we assume a 1~mm mylar layer as the only shielding material. The simulation's setup of parallel planar shielding layers hardly approximates the geometry of the experimental installation. Nonetheless, our simulations will yield a reasonable first estimate. For more precise determinations of the critical cross sections, the MC simulations would have to be generalized to more complicated simulation geometries, e.g.~using GEANT4~\cite{Agostinelli:2002hh}.

The projected modulation \textit{discovery} reach~(5$\sigma$) for these two experiments in the case of a light mediator are shown in Fig.~\ref{fig:satellite-balloon}, along with the combined low-mass direct-detection bounds.  
We see that a balloon-borne experiment could probe to larger cross sections by about two orders of magnitude above the current direct-detection constraints, while a satellite-borne experiment could probe an additional two orders of magnitude above a balloon-borne instrument.  

\subsubsection{Probing a subdominant component of dark matter interacting with an ultralight dark photon}\label{subsubsec:model}

We will now discuss whether a detector on a balloon or a satellite could probe open parameter space in a concrete DM model, where the open parameter space is unconstrained by current collider, astrophysical, or cosmological probes.  We have neglected this question above, as it is useful to present constraints and projections with as few model-dependent assumptions as possible.  
The following discussion is by no means complete, and additional work in this direction is warranted but beyond the scope of this paper.  

In general, DM that interacts through a heavy mediator is more constrained by collider and beam-dump searches than dark 
matter that interacts with a light mediator.  The reason is that in the light-mediator case, the direct-detection cross section, 
which scales as $\sim 1/q^4$, is enhanced by the low momentum transfer typical for a non-relativistic scattering event, while the high momentum transfer typical in relativistic collider or beam-dump events leads to a much smaller cross section.  
We will thus focus our discussion on the case where the DM interacts with a light mediator, although a careful analysis of 
the heavy mediator case is warranted.  

Examples of DM interacting with a light mediator include the case when the DM is millicharged (either with or without the existence of a massless dark photon mediator) or the DM interacts with a massive, but ultralight, dark photon.  
Both cases are strongly constrained at the large cross sections that could be probed by a balloon- or satellite-borne detector, and only a subdominant component of such DM is still viable.  
CMB observations set stringent bounds and limit the fractional DM abundance to be below 
about 0.4\%~\cite{Boddy:2018wzy}. For related work see e.g.~\cite{Chen:2002yh,Dvorkin:2013cea,Gluscevic:2017ywp,McDermott:2010pa,Slatyer:2015jla,Dolgov:2013una,Dubovsky:2003yn,Ali-Haimoud:2015pwa,Kadota:2016tqq,Boddy:2018wzy,Xu:2018efh,Barkana:2018qrx,Barkana:2018lgd,Munoz:2018pzp,Berlin:2018sjs,Dunsky:2018mqs}.  We will discuss other constraints further below. 

While a subdominant component of \textit{millicharged} DM, with or without a dark photon, is thus viable, it is possible that magnetic fields in the Galactic disk and supernovae will expel most of such DM from the Galactic disk~\cite{Chuzhoy:2008zy,McDermott:2010pa,Dunsky:2018mqs}.  Moreover, even if present in the Galactic disk, magnetic fields associated with the solar wind would decrease the DM flux observed on Earth substantially~\cite{Dunsky:2018mqs}.  While additional work analyzing the detectability of millicharged DM at large cross sections is warranted, it is clear that these considerations make it difficult to probe the high-cross-section region for millicharged DM with a balloon- or satellite-borne detector.  

On the other hand, a subdominant component of DM interacting with a \textit{massive}, albeit ultralight, dark photon, seems to be 
a viable model that could be explored at large cross sections with a balloon- or satellite-borne detector. 
For a massive dark-photon mediator, the effects from magnetic fields would be screened, with the screening length-scale corresponding roughly to the inverse of the mediator mass.  To screen the magnetic field effects on the scale of the solar system ($\sim 100$ AU) and larger, we need a dark photon mass larger than about $10^{-20}$~eV.  We then checked that a DM particle interacting with a dark photon of such small mass can penetrate the solar wind without much energy loss and deflection.  

\begin{figure*}
\centering
	\includegraphics[width=0.77\textwidth]{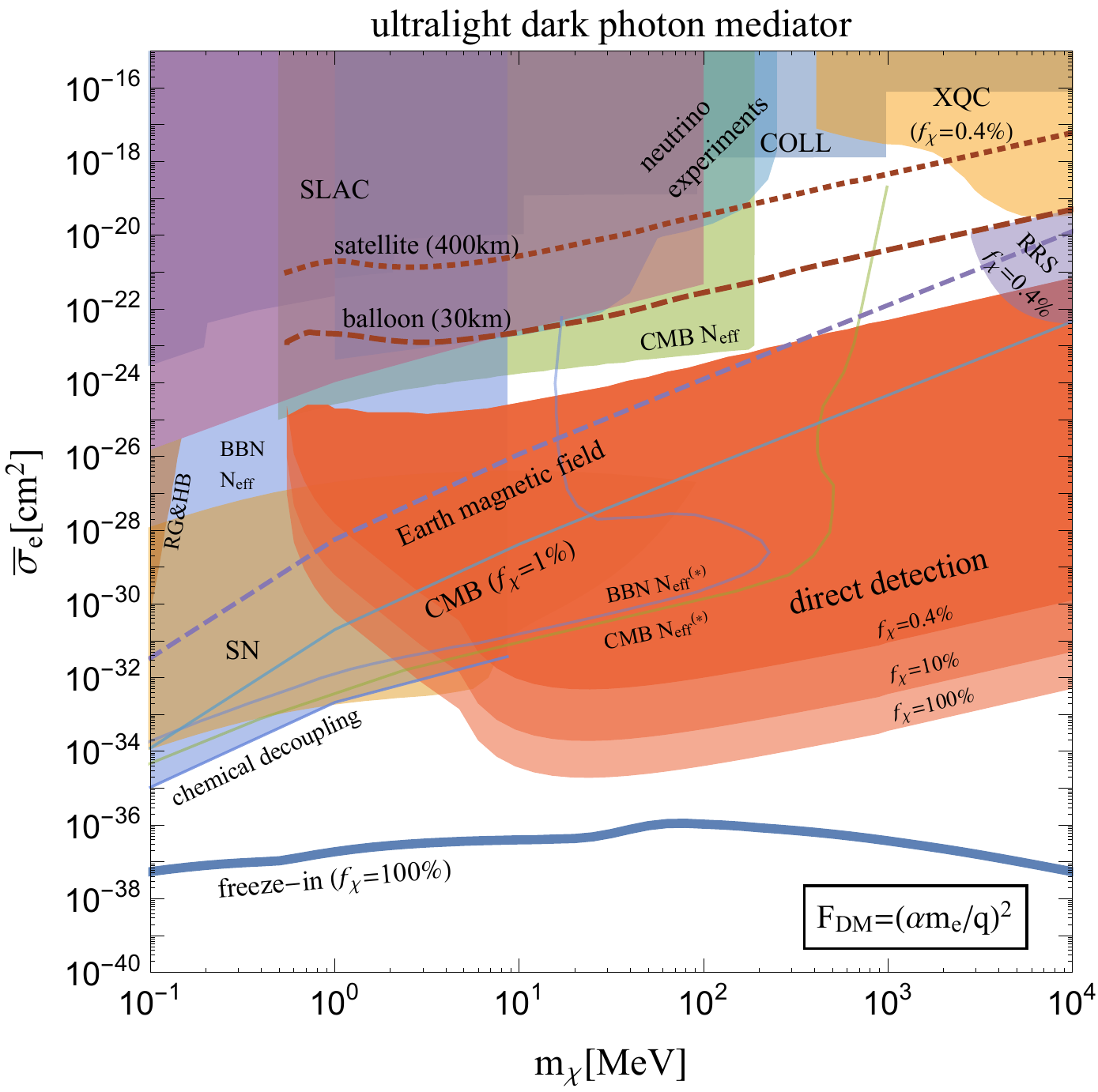}
	\caption{Discovery reach (thick red dashed lines) for a silicon dark matter detector with single-electron sensitivity on a balloon (satellite) assuming an exposure of 1 gram-hour (0.1 gram-month) and $10^6$ ($10^9$) background events, together with constraints on dark matter interacting with a massive, ultralight dark photon.  Also shown are cooling constraints from supernovae 1987A (brown, ``SN'')~\cite{Chang:2018rso}, 
	as well as Red-Giant and Horizontal-Branch stars~(brown, ``RG\&HB'')~\cite{Vogel:2013raa}; constraints from measurements of the number of relativistic degrees of freedom from the CMB (light green, ``CMB N$_{\rm eff}$'') and BBN (blue, ``BBN N$_{\rm eff}$'')~\cite{Creque-Sarbinowski:2019mcm,Munoz:2018pzp}, and from searches for milli-charged particles at SLAC (purple, ``SLAC'')~\cite{Prinz:1998ua}, colliders (blue, ``COLL'')~\cite{Davidson:1991si,Vogel:2013raa}, and at LSND and MiniBooNE (green, ``neutrino experiments'')~\cite{Magill:2018tbb}; 
and the direct-detection constraints derived in this paper from SENSEI, CDMS-HVeV, XENON10, XENON100, and DarkSide-50 (combined into one red-shaded region, labelled ``direct detection''), as well as from RRS (purple) and XQC (light orange)~\cite{Mahdawi:2018euy}.  We also show for comparison the ``freeze-in'' line along which DM obtains the correct relic density in this model~\cite{Essig:2011nj,Chu:2011be}.  The region at high cross sections is unconstrained from CMB measurements if this DM particle only makes up a subdominant component ($f_\chi \lesssim 0.4\%$) of the total observed DM abundance~\cite{Boddy:2018wzy}; for comparison, we show the CMB constraint for a fractional abundance of $f_\chi=1\%$ (blue line, ``CMB ($f_\chi=1\%$)''), in which case the entire region at high cross section is disfavored~\cite{Kovetz:2018zan}.  For the N$_{\rm eff}$ constraints from the CMB and BBN, we assume that the dark gauge coupling is sufficiently small to avoid the production of dark photons at early times; for large values of the dark gauge coupling, the bounds would be given by the thin blue and green lines (``BBN N$_{\rm eff^{(*)}}$'' and ``CMB N$_{\rm eff^{(*)}}$'')~\cite{Vogel:2013raa}.  
	}
	\label{fig: summary}
\end{figure*}

Fig.~\ref{fig: summary} shows a summary of the open parameter space for a DM particle interacting with a massive, ultralight dark photon mediator, and the potential reach of a balloon- and satellite-borne detector.  
We assume $f_\chi < 0.4\%$ to avoid bounds from the CMB as discussed above~\cite{Boddy:2018wzy}, but for comparison we show 
the CMB bound for $f_\chi = 1\%$ with a thin, blue line labelled ``CMB ($f_\chi=1\%$)''~\cite{Kovetz:2018zan}. 
We show several constraints that are independent of the DM abundance, including 
those from supernova cooling~\cite{Chang:2018rso}, stellar (red-giant and horizontal-branch-star) cooling~\cite{Vogel:2013raa}, and collider as well as proton-beam-dump searches for milli-charged particles~\cite{Davidson:1991si,Prinz:1998ua,Vogel:2013raa,Magill:2018tbb}.  
Also shown with an orange shaded region labelled ``direct detection'' are the combined direct-detection bounds derived in this paper from the two SENSEI prototype runs, CDMS-HVeV, XENON10, XENON100, and DarkSide-50 for fractional DM abundances of 
$f_\chi=0.4\%$, $10\%$, and $100\%$. The upper boundaries are not very sensitive to the precise DM abundance (as discussed in Sec.~\ref{subsubsec:exposure}). 

We also show constraints from the XQC experiment~\cite{Erickcek:2007jv} and RRS~\cite{Rich:1987st}.  
Constraints from conventional nuclear recoil experiments, such as the CRESST 2017 surface run~\cite{Angloher:2017sxg}, would be 
expected to be contained inside the region labelled ``direct detection''. 
XQC is a rocket-based experiment, and it therefore does exclude parameter space above the SENSEI and CDMS-HVeV constraints.  
We show in Fig.~\ref{fig: summary} the constraint for XQC, where we have naively rescaled to $f_\chi = 0.4\%$ the most conservative XQC constraints for $f_\chi = 1\%$ taken from~\cite{Mahdawi:2018euy}.  
At even higher masses ($m_{\chi} \geq 1$~GeV), we use data from a high-altitude nuclear recoil experiments labelled ``RRS'' to constrain the open parameter space between the upper boundary of the direct-detection experiments done on the surface and the lower boundary of the XQC experiment\footnote{With RRS we denote the DM~constraints derived by Rich, Rocchia, and Spiro from a 1977 balloon-borne experiment with a 0.5 gram silicon target~\cite{Rich:1987st}. From the reported atmospheric column density of 4.5$\text{gram/cm}^2$, we find that the maximum altitude must have been well above 30km, which is why we conservatively set the upper boundary of the constraints to the discovery reach line of our proposed balloon experiment. We estimate the lower boundary of the constraints by simply matching the event rate of the first bin of Fig.~1 in~\cite{Rich:1987st}. Compared to~\cite{Mack:2007xj,Hooper:2018bfw}, our resulting constraints are conservative.}.  While there is some uncertainty on the altitude of the RRS balloon flight, it closes the gap between XQC and the 
direct-detection experiments on the surface. 

There are also constraints from the CMB and BBN on the effective number of relativistic degrees of freedom, ${\rm N}_{\rm eff}$~\cite{Creque-Sarbinowski:2019mcm,Munoz:2018pzp}, labelled in Fig.~\ref{fig: summary} as ``CMB ${\rm N}_{\rm eff}$'' and ``BBN ${\rm N}_{\rm eff}$'', respectively.  
The presence of both the DM and the dark photon impacts the precise bound.  
The dark photon contribution to the $\text{N}_{\text{eff}}$ limit depends on the abundance of dark photons during the time of the CMB or BBN, which depends on the value of the dark photon coupling constant, $g_D$, and the kinetic mixing parameter, $\epsilon$.  
For $\epsilon \leq 2 \times 10^{-5} (\frac{m_{\chi}}{\text{MeV}})^{1/4}$, the $\text{N}_{\text{eff}}$ constraints can be significantly reduced if 
$g_D$ is sufficiently small, $g_D \leq 6 \times 10^{-6} (\frac{m_{\chi}}{\text{MeV}})^{1/4}$~\cite{Munoz:2018pzp}.   
For these low values of $g_D$, the production of the dark photons can be suppressed until the time of the CMB and hence the strong BBN and CMB bounds on $\text{N}_{\text{eff}}$ computed in~\cite{Vogel:2013raa} can be mostly evaded. Nonetheless, we show also  these stronger bounds applicable for larger values of $g_D$ with a thin solid blue and green line in Fig.~\ref{fig: summary}, labelled by ``BBN N$_{\rm eff^{(*)}}$'' and ``CMB N$_{\rm eff^{(*)}}$'' respectively.  The resulting region excluded from the CMB is shaded green in Fig.~\ref{fig: summary} and taken from~\cite{Munoz:2018pzp}.  The resulting BBN bound, shaded blue in Fig.~\ref{fig: summary}, comes from 
the DM population present during the time of BBN.  This BBN bound on ${\rm N}_{\rm eff}$ can be evaded if the DM is non-relativistic before neutrino decoupling or if the DM is not in chemical or in kinetic equilibrium with the SM around the time of neutrino decoupling.   The parameter space below which the DM chemically decouples from the SM is given by Eq.~(5) in~\cite{Creque-Sarbinowski:2019mcm} and shown in Fig.~\ref{fig: summary} as a thin blue boundary of the blue shaded region.  
The mass range above which the DM is sufficiently non-relativistic around the epoch of neutrino decoupling is given by $m_\chi \leq 8.62$ MeV~\cite{Creque-Sarbinowski:2019mcm} (assuming the DM is a Dirac fermion).  

To summarize, a satellite- or balloon-borne detector could probe open parameter space of a subdominant component of DM (abundance $\lesssim 0.4\%$) interacting with a dark photon of mass $\gtrsim 10^{-20}$~eV.  For couplings $\epsilon \leq 2 \times 10^{-5} (\frac{m_{\chi}}{\text{MeV}})^{1/4}$ and $g_D \leq 6 \times 10^{-6} (\frac{m_{\chi}}{\text{MeV}})^{1/4}$ the size of the unconstrained parameter space is much larger. The  satellite- or balloon-borne detector could consist of, for example, a Skipper-CCD.  However, we note that even an ``ordinary'' (non-Skipper) CCD (with higher noise, and therefore a higher mass threshold) could probe much of the open parameter space of this particular model, although we leave a detailed analysis of this to future work.  

Additional work is needed to fully analyze the direct-detection signal in this model.  In particular, the local DM density and velocity distribution may be affected by interactions with expanding supernovae remnants and interactions of the DM with nuclei and ions in the galaxy.  
In addition, 
Earth's magnetic field may affect the motion of the DM near Earth.  In particular, if the screening length of the dark photon is approximately more than the radius of the Earth (i.e., if the dark photon mass is less than $\sim 10^{-14}$ eV), particles follow the magnetic field lines and 
preferentially come close to the Earth near its magnetic poles.  
We find the critical cross section above which the particle flux near the magnetic poles is more than twice the flux for a particle unaffected by 
Earth's magnetic fields, and show this cross section with a purple dashed line labelled ``Earth magnetic field'' in Fig.~\ref{fig: summary}.  
For sufficiently high cross sections, 
the flux of DM particles near the equator can diminish completely, while the flux near the magnetic poles is enhanced significantly.  
The correct positioning and orbit of a balloon- or satellite-borne detector is therefore important, and also presents another handle on the DM signal.         

\section{Conclusions}\label{s:conclusions}

In this work, we investigated how terrestrial effects limit the sensitivity to dark matter that has large interaction strengths with ordinary matter for direct-detection experiments searching for sub-GeV DM through electron recoils. We considered a model in which DM interacts with a dark photon, including both cases of a heavy and ultralight mediator, as well as DM that interacts with an electric dipole moment, and DM that interacts only with electrons. For these models, we determined the critical DM-electron cross section above which the direct-detection experiments lose sensitivity because of interactions in the overburden. We re-analyzed the data from SENSEI, CDMS-HVeV, XENON10, XENON100, and DarkSide-50 to calculate the corresponding excluded band of cross sections, and also determined projected sensitivities for future experiments.  
We derived, for example, the sensitivity of the proposed SENSEI and DAMIC-M experiments taking into account the terrestrial effects, see Fig.~\ref{fig:ProjectionSenseiDamic}.  We find that their sensitivity at small cross sections is largely unaffected by terrestrial effects.

We considered the terrestrial effect arising from interactions between DM and nuclei and electrons in the Earth's atmosphere and crust.  
We considered DM scattering elastically off electrons and nuclei, as well as inelastic scatters off electrons leading to atomic ionization and electronic excitations in semiconductors.  In the case of elastic DM-nuclei scatterings, we used MC techniques to estimate the stopping effect. On the other hand, we used analytic methods to estimate the stopping power of elastic and inelastic scatterings of DM with electrons in the overburden. We found that in the range of the masses considered in the paper and for DM velocities typical of those in the Milky-Way halo, nuclear stopping dominates the stopping effect from electrons (see Fig.~\ref{fig:estopping}). Therefore, when the DM couples to the nuclei, the upper limit on the DM-electron cross-section is determined by the stopping effect from nuclei, see solid lines in Figs.~\ref{fig:constraints} and~\ref{fig:ProjectionSenseiDamic}. In the case where DM couples exclusively to electrons, the stopping effect is determined by the much weaker electronic stopping effect, as seen in the dashed lines in Figs.~\ref{fig:constraints} and~\ref{fig:ProjectionSenseiDamic}. 

We also found that the terrestrial effect is largely insensitive to the exposure, since the number of events observed in the detector drops sharply near the critical cross section, see Fig.~\ref{fig:Projection_exposure}.  
On the other hand, a larger overburden of the experiment significantly decreases the critical cross section, see Fig.~\ref{fig:Projection_depth}. 
As the depth increases, the stopping effect increases and the critical cross section decreases. 

It is well known that terrestrial effects limit the sensitivity of direct detection experiments to dark matter with large interaction strengths. This motivates the use of a balloon or a satellite as the site of a direct-detection experiment. 
In that case, there is a significant shadowing effect from the Earth, and the signal would be characterized by a strong modulation. We calculated the modulated event rate in a silicon semiconductor experiment placed on a balloon over Pasadena, California and in the ISS, for a DM particle interacting with an ultralight dark photon, see Fig.~\ref{fig: modulation}.  

For DM interacting with the SM through an ultralight dark photon, a balloon- or satellite-borne experiment is further motivated as there is open parameter space above current constraints from direct-detection experiments done on the surface and below current collider and beam-dump experiments. The fractional DM density needs to be below about $0.4\%$ to avoid cosmological constraints, and more parameter space opens up if we assume the dark coupling constant to be sufficiently small to avoid production of dark photons in the early Universe, see Fig.~\ref{fig: summary} and Sec.~\ref{subsubsec:model} for more details. The inability of surface or underground experiments to probe such high cross sections could make a satellite- or balloon-borne detector crucial in constraining this region. Further work is needed, however, to understand the local DM density and velocity distribution at such high cross sections for this DM model, since these may be affected by DM interactions with supernova remnants and with nuclei and ions in the Galaxy.  We leave the detailed modelling of these effects to future work.

\acknowledgments

We thank John Beacom, Jae Hyeok Chang, Juan Estrada, Hunter Hall, Roni Harnik, Gordan Krnjaic, Nadav Outmezguine, Samuel McDermott, Matthew Pyle, Diego Redigolo, Javier Tiffenberg, Yu-Dai Tsai, Tomer Volansky, and Tien-Tien Yu for useful discussions. 
RE and MS are supported by DoE Grant DE-SC0017938.  RE also acknowledges support from the US-Israel Binational Science Foundation 
under Grant No.~2016153. 
TE and CK are partially funded by the Danish National Research Foundation, grant number DNRF90, and by the Danish Council for Independent Research, grant number DFF 4181-00055. 
TE was supported by the Knut and Alice Wallenberg Foundation (PI, Jan Conrad). 
Computation/simulation for the work described in this paper was supported by the DeIC National HPC Centre, SDU. 
TE thanks the C.N.~Yang Institute for Theoretical Physics for the hospitality during a visit where parts of this work were developed.
 
\appendix

\section{Rare event simulation}
\label{app:splitting}
For large cross sections, the shielding's stopping power reflects or stops the incoming DM flux, and DM particles rarely reach the detector while still being detectable. The brute force MC simulation of these rare events is computationally extremely expensive, inefficient, and in many cases practically impossible. More advanced MC methods such as variation reduction techniques can be of great benefit. In a previous work~\cite{Emken:2018run}, Importance Sampling (IS) was used to speed up the simulations of DM particles in matter, as first proposed in~\cite{Mahdawi:2017cxz}. Therein, the authors modify the simulations' probability density functions~(PDFs) by
\begin{enumerate}
	\item increasing the mean free path.
	\item favouring forward scattering.
\end{enumerate}
Both biases increase the probability of a particle to reach and trigger the underground detector. More crucially, the changes of the distributions mimic the distribution of the successful rare events in cases where the DM particle typically loses a significant fraction of its kinetic energy in a single scattering. Therefore, the method works best for DM with mass $\mathcal{O}$(10~GeV) and contact interactions. In contrast, if a single scattering only causes a tiny relative loss of kinetic energy, such as e.g. for very light or very heavy DM, in particular with ultralight mediators, detectable particles can have scattered hundreds or thousands of times before reaching the detector. In these cases, the two IS~modifications no longer imitate the successful particles, and the results obtained with~IS become unstable and unreliable.

\subsection{Adaptive Geometric Importance Splitting}
A well-studied alternative rare event simulation method is Geometric Importance Splitting~(GIS)\footnote{For an introduction to splitting techniques, we refer to~\cite{Haghighat2016}.}, first described by Kahn and Harris in the context of neutron transport and shielding~\cite{Kahn1951}. As opposed to~IS, the use of~GIS does not introduce a bias in the underlying PDFs, and the sampling of the random variables is unchanged. Instead, ``important'' particles are being split into multiple copies, each of which is propagated further independently. Furthermore, ``unimportant'' particles have a chance to be eliminated. The physical intuition behind this method is the fundamental notion that we simulate particle packages, and not individual particles.

The central object of GIS is the so-called importance function $I:\mathbb{R}^3\rightarrow \mathbb{R}$, which defines ``important'' particles. In our case the importance function is a function of the particle's underground depth only, which increases the closer the particle approaches the detector depth.

We assume a particle has a statistical weight $w_i$, which would correspond to the particle package's size, and importance $I_i$. After it scatters at depth $z$, we compare its previous importance to the new value $I_{i+1}\equiv I(z)$. If
\begin{align}
	\nu &\equiv \frac{I_{i+1}}{I_i} > 1\, ,
	\intertext{we split the particle into a number of $n$ copies, each of which is assigned the weight}
	w_{i+1}&\equiv \frac{w_i}{n}\, .
\intertext{The number of copies is given by,}
n&=\begin{cases}
	\nu\, ,\quad &\text{if }\nu\in\mathbb{N}\, ,\\
	\lfloor \nu\rfloor \, ,\quad &\text{if }\nu\notin\mathbb{N} \land \xi \geq \Delta\, ,\\
	\lfloor \nu\rfloor+1\, ,\quad &\text{if }\nu\notin\mathbb{N} \land \xi<\Delta \, ,
\end{cases}
\label{eq:giscases}
\end{align}
where $\Delta\equiv \nu-\lfloor \nu \rfloor$ is the non-integer part of $\nu$ and $\xi\in(0,1)$ is a uniformly distributed random number. This way, the expectation value of $n$ for non-integer $\nu$ is $\langle n\rangle = \Delta (\lfloor\nu\rfloor+1) + (1-\Delta)\lfloor \nu\rfloor = \nu$.

The counterpart to particle splitting is called \textit{Russian Roulette}. If a particle becomes less important, i.e. 
\begin{align}
	\nu &= \frac{I_{i+1}}{I_i} < 1\, ,
\end{align}
there is a probability $p_{\rm kill}= 1-\nu$ that the particle is eliminated and the simulation stops. If the particle survives, it is assigned the new weight
\begin{align}
	w_{i+1} = \frac{w_i}{1-p_{\rm kill}} = \frac{w_i}{\nu} >w_i\, .
\end{align}
This ensures that the expectation value of the new weight remains unchanged, as $\langle w_{i+1}\rangle = p_{\rm kill}\cdot 0 + (1-p_{\rm kill})\cdot w_{i+1} = w_i$.  Note that Russian roulette is nothing but a special case of Eq.~\eqref{eq:giscases}. Overall, this weighting procedure ensures that splitting and Russian Roulette do not introduce biases into the MC estimates.
\begin{figure}
	\centering
	\includegraphics[width=0.49\textwidth]{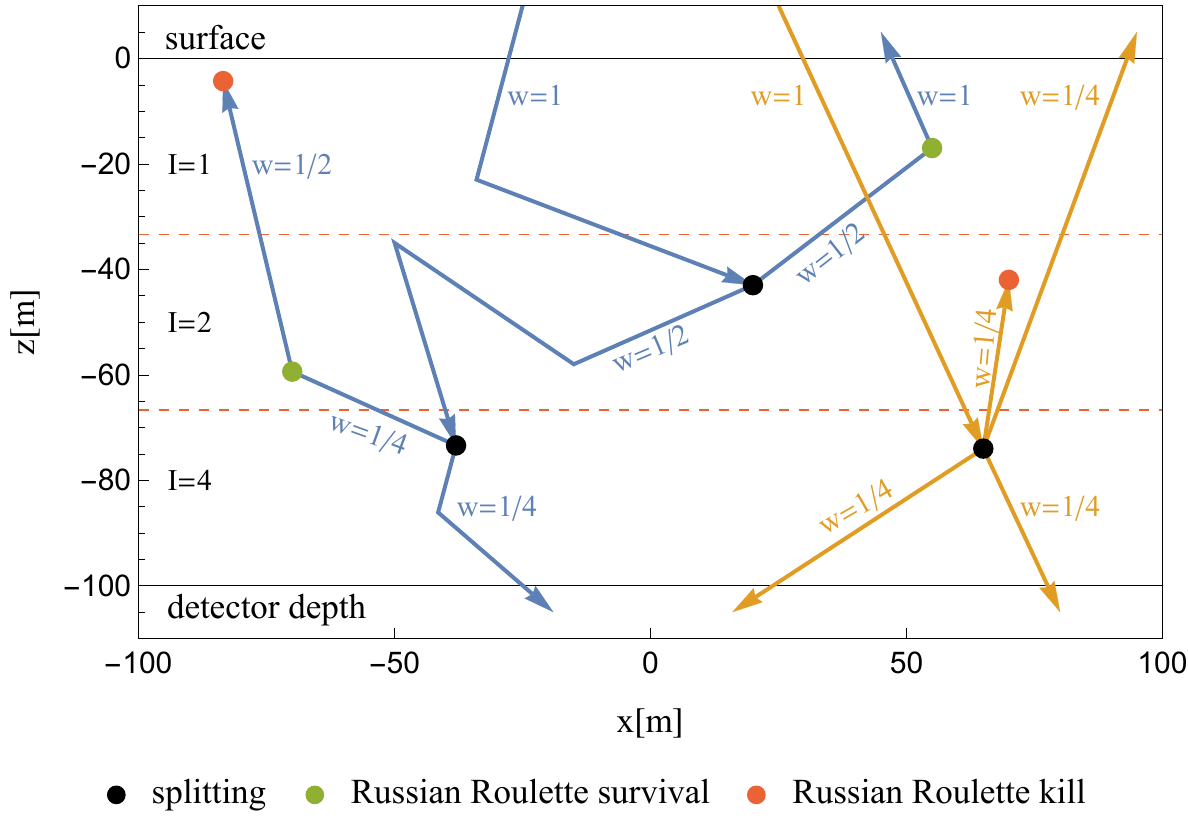}
	\caption{Illustration of GIS for two DM trajectories with three importance domains of equal size. The particle can split into either 2 or 4 copies. The weight development along the paths is shown as well.}
	\label{fig:gis}
\end{figure}

\subsection{The Importance Function and Adaptive GIS}
In the case of IS the central challenge was to find the optimal modification of the simulation's PDFs, whereas the main problem for GIS is to find a good importance function, which quantifies how close we are to the detector depth $d$. One possibility is to divide the shielding layers into $N_I$ domains of constant importance. We define a sequence of planar splitting surfaces with depths $0>l_1>l_2>...>l_{N_I-1}>d$, which define layers of increasing importance. By assigning domain $k$, for example, the importance $I_k=N_{\rm splits}^{k-1}$, we assure that a particle from domain $k$ reaching $k+1$ splits into $N_{\rm splits}$ copies on average, as~$N_{\rm splits}$ does not have to be an integer. Finally, there are two questions that need to be addressed. What is the optimal number $N_I$ of importance domains, and given an answer, how are the locations $l_k$ of the splitting surfaces determined?

The number of importance domains should be determined adaptively. If $N_I$ is chosen too large, a single particle might pass multiple splitting surfaces and split into a large number of copies. For example, if we were to use $N_I=10$, and $N_{\rm splits}=3$ in the single scattering regime, a particle which scatters in domain 9 for the first time splits into $3^8=6561$ copies. If a large fraction of these copies reach the detector depth, the final data set will be highly correlated.

For hard scatterings one might just set $N_I \sim d / \lambda$, using the mean free path $\lambda$. However, if a single scattering causes only a small relative loss of energy, this number is no longer a good choice. Instead, we determine $N_I$ via the average integrated stopping power using eq.~\eqref{eq:stopping},
\begin{align}
	\langle\Delta E_\chi\rangle &\equiv \int\limits_{0}^d \dd x \,S_n(m_\chi,\sigma_p,\langle v_\chi\rangle) = \sum_{l=1}^{N_{\rm layers}} t_l S_n^l(m_\chi,\sigma_p,\langle v_\chi\rangle)\, .
\end{align}
In the second step, we used that each physical shielding layer is defined by a constant density and composition, and hence stopping power. Furthermore, $t_l$ is the thickness of shielding layer $l$, and $\langle v_\chi\rangle = \int_{v_{\rm cutoff}}^{v_{\rm max}}\dd v\; v f(v)$ is the average initial speed of the simulated DM particles. An adaptive number of importance domains is then
\begin{align}
	N_I = \left\lceil\kappa\cdot \frac{\langle\Delta E_\chi\rangle}{\frac{m_\chi}{2}\left(\langle v_\chi \rangle^2-v_{\rm cutoff}^2\right)}\right\rceil\, ,
\end{align}
where $\kappa$ is a parameter that can be freely adjusted depending on how fast the number of domains is supposed to increase. Next, we have to determine the location of the splitting surfaces at depth $l_1,...,l_{N_I-1}$. Here it is of great advantage, if all domains are of equal integrated stopping power, i.e.,  
\begin{align}
	\langle \Delta E_\chi\rangle_1=\langle \Delta E_\chi\rangle_2=\ldots=\langle \Delta E_\chi\rangle_{N_I}=\frac{\langle \Delta E_\chi\rangle}{N_I}\, ,
\end{align}
where~$\langle\Delta E_\chi\rangle_i \equiv \int_{l_{i-1}}^{l_i} \dd x \,S_n$ is the average energy loss in the~$i$-th importance domain. This fixes the locations and ensures that we do not place too many of the boundaries in regions of relatively low stopping power.

Compared to non-GIS MC simulations this method yields speed-ups of up to two orders of magnitude. The results obtained this way have been verified for multiple examples by direct comparison to brute force MC simulations.

\section{DM-Electron scattering experiments}
\label{app:experiments}

\subsection{XENON10 and XENON100}

The observation of `S2-only' events in XENON10~\cite{Angle:2011th} and XENON100~\cite{Aprile:2016wwo} data were used to derive bounds on the DM-electron scattering cross section in~\cite{Essig:2012yx,Essig:2017kqs}. The ionization rate and spectrum are given in Eq.~\eqref{noble}. For the details on the atomic shells, the secondary electrons, and the transformation from the theoretical spectrum to the observed PE spectrum~$\frac{\dd R}{\dd \text{PE}}$, we refer to~\cite{Essig:2012yx,Essig:2017kqs}.

The data XENON10~(XENON100) corresponds to an exposure of 15 kg-days (30 kg-years). Both experiments were located at Gran Sasso beneath 1400m of rock. The binned numbers of events in terms of the photoelectrons (PE) are listed in Table~\ref{tab: xenon}.

\begin{table}
	\centering
	\begin{tabular}{|ll|c|ll|}
	\cline{1-2}\cline{4-5}
	\textbf{XENON10}			&			&\phantom{space}&\textbf{XENON100}	&			\\
	bin~[PE]				&events		&&bin~[PE]				&events		\\
	\cline{1-2}\cline{4-5}
	$\text{[}$14,41)			&126			&&$\text{[}$80,90)		&794			\\
	$\text{[}$41,68)			&60			&&$\text{[}$90,110)		&1218			\\
	$\text{[}$68,95)			&12			&&$\text{[}$110,130)	&924			\\
	$\text{[}$95,122)		&3			&&$\text{[}$130,150)	&776			\\
	$\text{[}$122,149)		&2			&&$\text{[}$150,170)	&669			\\
	$\text{[}$149,176)		&0			&&$\text{[}$170,190)	&630			\\
	$\text{[}$176,203)		&2			&&$\text{[}$190,210)	&528			\\
	\cline{1-2}\cline{4-5}
	\end{tabular}
	\caption{Number of events observed in bins of photoelectrons (PE) for XENON10 (left) and XENON100 (right).}
	\label{tab: xenon}
\end{table}

The total efficiency for XENON10 is the product of a flat cut efficiency (92\%)~\cite{Angle:2011th} and the trigger efficiency taken from Fig.~1 of~\cite{Essig:2012yx}. For XENON100, the respective efficiencies can be found in Fig.~3 of~\cite{Aprile:2016wwo}.  Finally, the bound on the cross section for a given DM~mass is found using Poisson statistics independently in each bin. For XENON100, the first three bins yield the strongest constraints.

\subsection{SENSEI and SuperCDMS}
In this paper, we present constraints from three sets of data obtained with a silicon semiconductor target and with sensitivity to single electron excitations.  The event rates and spectra are computed with Eq.~\eqref{semiconductor}~\cite{Essig:2015cda}. Two of the data sets, namely from the SENSEI protoSENSEI@surface run~\cite{Crisler:2018gci} and the CDMS-HVeV~\cite{Agnese:2018col} run, were obtained on the surface, and are thus ideal to probe strong DM-electron interactions. Most recently, the SENSEI collaboration presented results for protoSENSEI@MINOS, a prototype skipper-CCD detector set up in the MINOS cavern at Fermilab with a relatively shallow underground depth of 107~m~\cite{Abramoff:2019dfb}.

Concerning protoSENSEI@surface, the exposure was $\sim$0.02 gram-days. The observed event numbers can be taken from Table~I in~\cite{Crisler:2018gci} or from Table~\ref{tab:sensei signals}. This experiment was performed inside the Silicon Detector Facility at Fermilab, and was only shielded by the atmosphere and a few cm of concrete roof, which we can neglect. 

\begin{table}[h!]
\centering
\begin{tabular}{l|ll|c|ll|c|}
	\cline{2-3}\cline{5-6}
			&\multicolumn{2}{l|}{\textbf{protoSENSEI@surface}}		&&\multicolumn{2}{l|}{\textbf{CDMS-HVeV}}	\\
	$n_e$\phantom{}	&efficiency		&events		&&efficiency		&events		\\
	\cline{2-3}\cline{5-6}
	1				&0.668			&140302		&&0.88			&$\sim$53000	\\
	2				&0.41			&4676		&&0.91			&$\sim$400	\\	
	3				&0.32			&131		&&0.91			&$\sim$74	\\
	4				&0.27			&1			&&0.91			&$\sim$18	\\
	5				&0.24			&0			&&0.91			&$\sim$7	\\	
	6				&--			&--				&&0.91			&$\sim$14	\\
	\cline{2-3}\cline{5-6}
\end{tabular}
	\caption{Efficiencies and numbers of observed signals in the electron bins for protoSENSEI@surface and SuperCDMS (`CDMS-HVeV').}
	\label{tab:sensei signals}
\end{table}

The CDMS-HVeV results by SuperCDMS are based on a surface run with an exposure of 0.487~gram-days. The efficiencies and signal number are listed in the center of Table~\ref{tab:sensei signals}. The efficiencies were taken from Fig.~3 of~\cite{Agnese:2018col}, whereas the event numbers were estimated on the basis of the histogram in the same figure. While the official constraints were obtained with Yellin's optimum interval method, we use Poisson statistics for each bin. Furthermore, the official analysis considered only data from within 2$\sigma$ of the electron peaks with $\sigma=$0.07~electron-hole pairs. We take over this procedure by using a flat efficiency factor $\epsilon\approx 0.9545$ corresponding to the $2\sigma$.

Regarding the shielding, we take the Earth's atmosphere into account as well as the~60~cm of concrete. The concrete is modeled as a layer with~$\rho=2.4\text{ g cm}^{-3}$, whose nuclear composition is listed in Table~\ref{tab:concrete}. 
\begin{table}[h!]
\centering
	\begin{tabular}{|ll|}
	\hline
		Element			&fraction~[wt\%]\\
	\hline
	\isotope{H}{1}		&0.33		\\
	\isotope{O}{16}		&52.28		\\
	\isotope{Na}{23}		&0.02		\\
	\isotope{Mg}{24}		&0.10			\\
	\isotope{Al}{27}		&0.33			\\
	\isotope{Si}{28}		&40.85			\\
	\isotope{S}{32}		&0.16			\\
	\isotope{Cl}{35}		&0.01			\\
	\isotope{K}{39}		&0.06			\\
	\isotope{Ca}{40}		&5.59			\\
	\isotope{Fe}{56}		&0.27			\\
	\hline
	{\bf Total}			&{\bf 100.0}			\\	
	\hline
	\end{tabular}
	\caption{Composition of concrete \cite{Piotrowski2012}.}
	\label{tab:concrete}
\end{table}

For the protoSENSEI@MINOS data, we use the efficiencies and event numbers as listed on the right hand side of Table~I of~\cite{Abramoff:2019dfb}. These events were observed during a combined exposure of 0.246~gram-days. The constraints are obtained as limits on the observed signal rate as described in~\cite{Abramoff:2019dfb}. 

For all of these three data sets, we compute the limits using Poisson statistics independently for each bin. The most constraining bin sets the overall bound.

\subsection{DarkSide-50}
The DarkSide collaboration presented constraints on sub-GeV~DM based on data collected by the DarkSide-50 argon detector~\cite{Agnes:2018oej}. The data has an exposure of 6786~kg~days and a threshold of three electrons. Our own derivation of the constraints using this data yielded weaker constraints than those presented in~\cite{Agnes:2018oej}, most likely due to a differing derivation of the ionization form factors for argon.  We show in Fig.~\ref{fig:constraints} the constraint for contact and long-range interactions taken directly from Fig.~4 of~\cite{Agnes:2018oej}, but use our own derivation of the result for $F_{\rm DM}(q)\propto 1/q$.   For the upper boundaries, we also use our own derivation, but this has negligible impact on calculating the critical cross section as is seen in, for example, Fig.~\ref{fig:Projection_exposure}. 

\section{Derivation of the electronic stopping power in Silicon}\label{app:derivation}

Using Eq.~(A.12) in~\cite{Essig:2015cda}, we can write the stopping power of an electronic transition from initial energy level 1 to final energy level 2 for a DM particle traveling with a velocity $\textbf{v}$,
\begin{multline}
\left.S_e\right|_{1\rightarrow 2}=\frac{n_{\text{T}} \overline{\sigma}_{e}}{\mu_{\chi e}^2 v_{\rm rel}}\int \frac{\dd^3 q}{4 \pi} (\Delta E_{1 \rightarrow 2}) \delta(\Delta E_{1 \rightarrow 2} + \frac{q^2}{2 m_{\chi}}- qv \cos{\theta_{qv}})\\ \times \left|F_{\rm DM}(q)\right|^2 \left|f_{12}(q)\right|^2 \Theta[v-v_{\rm min}(\Delta E_{1 \rightarrow 2},q)]\,,  
\end{multline}
where we have inserted a factor of $\Delta E_{1 \rightarrow 2}$ corresponding to the energy loss of the DM~particle in the integral and multiplied by the number density of the overburden $n_{\text{T}}$. As the DM particles have a velocity distribution, we can average the stopping power over the angular distribution to get the average stopping power for a DM particle with speed $v$ as,
\begin{multline}
\left.\langle S_e \rangle\right|_{1\rightarrow 2}= \frac{n_{\text{T}} \overline{\sigma}_{e}}{\mu_{\chi e}^2 v_{\rm rel}}\int \frac{\dd^3 q}{4 \pi}\times\frac{1}{2}\int_{-1}^{1} \dd \cos{\theta_{qv}} (\Delta E_{1 \rightarrow 2}) \delta(\Delta E_{1 \rightarrow 2} + \frac{q^2}{2 m_{\chi}}- qv \cos{\theta_{qv}})\\ \times \left|F_{\rm DM}(q)\right|^2 |f_{12}(q)|^2\Theta[v-v_{\rm min}(\Delta E_{1 \rightarrow 2},q)] \,,  
\end{multline}
where we assume an isotropic velocity distribution. The angular integral can be used to remove the energy delta function. This gives,
\begin{multline}\label{eq:general}
\langle S_e \rangle |_{1\rightarrow 2}= \frac{n_{\text{T}} \overline{\sigma}_{e}}{\mu_{\chi e}^2 v_{\rm rel}}\int \frac{\dd^3 q}{8 \pi} (\Delta E_{1 \rightarrow 2})\times \frac{1}{qv}\times |F_{\rm DM}(q)|^2 |f_{12}(q)|^2 \Theta[v-v_{\rm min}(\Delta E_{1 \rightarrow 2},q)].  
\end{multline}
For a semiconductor crystal, the transition factor $|f_{12}(q)|^2$ is replaced by a form factor $|f_{ i\vec{k} \rightarrow i'\vec{k'}}|^2$ to excite from the valence level $\{ i\vec{k}\}$ to a conduction level $\{ i'\vec{k'}\}$ and is defined in Eq.~(A.25) of~\cite{Essig:2015cda} as 
\begin{equation}\label{eq:def}
\left|f_{ i\vec{k} \rightarrow i'\vec{k'}}\right|^2 = \sum_{\vec{G'}}\frac{(2\pi)^3 \delta^3(\vec{q}-(\vec{k'}+\vec{G'}-\vec{k}))}{V}\left|\sum_{\vec{G}}u_{i'}^*(\vec{k'}+\vec{G}+\vec{G'})u_i(\vec{k}+\vec{g})\right|^2\,.
\end{equation}
Replacing $n_{\text{T}}$ by $n_{\text{cell}}$, which is the number density of the unit cells in the crystal, and inserting Eq.~(\ref{eq:def}) into Eq.~(\ref{eq:general}), we get  
\begin{equation}
\langle S_e \rangle |_{i\vec{k}\rightarrow i'\vec{k'}}= \frac{n_{\text{cell}} \overline{\sigma}_{e}\pi^2}{\mu_{\chi e}^2 v_{\rm rel}V}\sum_{\vec{G'}} (\Delta E_{i \rightarrow i'})\Theta[v-v_{\rm min}(\Delta E_{1 \rightarrow 2},q)]\times \frac{1}{qv}\times |F_{\rm DM}(q)|^2 |f_{[i\vec{k},i'\vec{k'},\vec{G'}]}|^2 \bigg|_{q=|\vec{k'}+\vec{G'}-\vec{k}|}\,,  
\end{equation}
where $|f_{[i\vec{k},i'\vec{k'},\vec{G'}]}|^2$ is the term in the square of the absolute value in Eq.~(\ref{eq:def}). Summing over the initial and final states, we get the total stopping power as,
\begin{multline}
\langle S_e \rangle= \frac{n_{\text{cell}}V_{\text{cell}} \overline{\sigma}_{e}2\pi^2}{\mu_{\chi e}^2 v_{\rm rel}}\sum_{ii'}\int_{BZ} \frac{\dd^3 k \dd^3 k'}{(2\pi)^6}\sum_{\vec{G'}} (\Delta E_{i \rightarrow i'})\Theta[v-v_{\rm min}(\Delta E_{1 \rightarrow 2},q)]\\ \times \frac{1}{qv}\times |F_{\rm DM}(q)|^2 |f_{[i\vec{k},i'\vec{k'},\vec{G'}]}|^2 \bigg|_{q=|\vec{k'}+\vec{G'}-\vec{k}|}\,.
\end{multline}
Note that we have the volume of a unit cell $V_{\text{cell}}$ in this expression as compared to the total volume $V$ in Eq.~(A.30) in~\cite{Essig:2015cda} because the phase space for the initial energy level is just one unit cell that the dark matter particle is passing through. Inserting the energy and momentum delta functions, we get  
\begin{multline}
\langle S_e \rangle= \frac{n_{\text{cell}}V_{\text{cell}} \overline{\sigma}_{e}2\pi^2}{\mu_{\chi e}^2 v_{\rm rel}}\int \dd\text{ln} E_e \dd\text{ln} q \;(E_e)\Theta[v-v_{\rm min}(E_e,q)]\times \frac{1}{qv}\times |F_{\rm DM}(q)|^2 \\ \times \sum_{ii'}\int_{BZ} \frac{\dd^3 k \dd^3 k'}{(2\pi)^6}E_e \delta (E_e - E_{i'\vec{k'}}+E_{i\vec{k}})\sum_{\vec{G'}}q\delta{q-|\vec{k'}+\vec{G'}-\vec{k}|}|f_{[i\vec{k},i'\vec{k'},\vec{G'}]}|^2\,.
\end{multline}
This can be written as,
\begin{equation}
\langle S_e \rangle=n_{\rm cell} \bar{\sigma}_{e} \alpha \times \frac{m_{e}^{2}}{\mu_{\chi e}^{2}v v_{\rm rel}} \int \dd E_{e} E_{e} \int \frac{\dd q}{q^2} F_{\rm DM}(q)^2 \left| f_{\rm crystal} (q,E_e) \right|^2 \Theta[v-v_{\rm min}(E_{e},q)]\,,     
\end{equation}
where $\left| f_{\rm crystal} (q,E_e) \right|^2$ is as defined in Eq.~(A.33) in~\cite{Essig:2015cda},
\begin{multline}
\left| f_{\rm crystal} (q,E_e) \right|^2= \frac{2\pi^2(\alpha m_e^2 V_{\text{cell}})^{-1}}{E_e} \sum_{ii'}\int_{BZ} \frac{V_{\text{cell}}\dd^3 k }{(2\pi)^3}\frac{V_{\text{cell}}\dd^3 k' }{(2\pi)^3}\times \\ E_e \delta (E_e - E_{i'\vec{k'}}+E_{i\vec{k}})\sum_{\vec{G'}}q\delta{q-|\vec{k'}+\vec{G'}-\vec{k}|}|f_{[i\vec{k},i'\vec{k'},\vec{G'}]}|^2\,.
\end{multline}
This reproduces Eq.~(\ref{eq:ionizationstopping}).

\bibliographystyle{JHEP}
\bibliography{library}
\end{document}